\documentclass[iicol]{submission-style}

\usepackage[utf8]{inputenc}
\usepackage{CJKutf8}
\usepackage{amsmath}
\usepackage{amssymb}
\usepackage{amsfonts}
\usepackage{geometry}
\usepackage{braket}
\usepackage{graphicx}
\usepackage{float} 
\usepackage{booktabs}
\usepackage{array}
\usepackage{multirow}
\usepackage{caption}
\usepackage{subcaption}
\usepackage[numbers,sort&compress]{natbib}
\usepackage{ulem}
\usepackage{hyperref}
\usepackage{lineno}
\usepackage{xcolor}
\begin{document}
\begin{CJK}{UTF8}{gbsn}
	
	\title{Quantum-Classical Hybrid Computation of Electron Transfer in a Cryptochrome Protein via VQE-PDFT and Multiscale Modeling}
	
	\author[1]{\fnm{Yibo} \sur{Chen} (陈一博)}
	
	\author[2]{\fnm{Zirui} \sur{Sheng} (盛子瑞)}
	
	\author[2]{\fnm{Weitang} \sur{Li} (李维唐)}
	
	\author[1,3,4]{\fnm{Yong} \sur{Zhang} (张勇)}
	
	\author[1]{\fnm{Xun} \sur{Xu} (徐讯)}

	\author*[1]{\fnm{Jun-Han} \sur{Huang} (黄俊翰)}\email{huangjunhan@genomics.cn}
	
	\author*[3,4]{\fnm{Yuxiang} \sur{Li} (黎宇翔)}\email{liyuxiang@genomics.cn}
	
	\affil[1]{\orgdiv{State Key Laboratory of Genome and Multi-omics Technologies}, \orgname{BGI Research}, \orgaddress{\city{Shenzhen}, \postcode{518083}, \country{China}}}
	
	\affil[2]{\orgdiv{School of Science and Engineering}, \orgname{The Chinese University of Hong Kong}, \orgaddress{\city{Shenzhen}, \state{Guangdong}, \postcode{518172}, \country{P.R.China}}}
	
	\affil[3]{\orgname{BGI Research}, \orgaddress{\city{Wuhan 430047}, \country{China}}}
	
	\affil[4]{\orgdiv{Guangdong Bigdata Engineering Technology Research Center for Life Sciences}, \orgname{BGI Research}, \orgaddress{\city{Shenzhen 518083}, \country{China}}}
	
	\abstract{Accurate calculation of strongly correlated electronic systems requires proper treatment of both static and dynamic correlations, which remains challenging for conventional methods. To address this, we present VQE-PDFT, a quantum-classical hybrid framework that integrates variational quantum eigensolver with multiconfiguration pair-density functional theory (MC-PDFT). This framework strategically employs quantum circuits for multiconfigurational wavefunction representation while utilizing density functionals for correlation energy evaluation. The hybrid strategy maintains accurate treatment of static and dynamic correlations while reducing quantum resource requirements compared to highly expressive quantum algorithms. Benchmark validation, performed via noiseless quantum circuit simulator, on the Charge-Transfer dataset confirmed that VQE-PDFT achieved results comparable to conventional MC-PDFT. Building upon this, we developed shallow-depth hardware-efficient ansatz circuits and integrated them into a QM/MM multiscale architecture to enable applications in complex biological systems. This extended framework, when applied to electron transfer in the European robin cryptochrome protein ErCRY4 with noiseless simulations, yielded transfer rates that aligned well with experimental measurements. Finally, as a proof-of-concept hardware demonstration, we executed the reduced-density-matrix measurements for a single protein conformation on a 13-qubit superconducting device and showed the impact of noise through a comprehensive error analysis.}
	
	
	\maketitle
	
	\section{Introduction}\label{intro}
	
	Evaluating strongly correlated systems is one of the most fundamental challenges in modern science, where conventional theoretical approaches often struggle due to the intricate electronic coupling. The accurate description of these systems has led to major breakthroughs across multiple disciplines: in physics, it has driven the understanding of high-temperature superconductivity \cite{yanagisawa2019}; in chemistry, it has enabled the precise modeling of catalytic processes \cite{biz2021}; in biology, it has provided crucial insights into the mechanisms of nitrogen fixation \cite{reiher2017} and avian magnetoreception \cite{zhang2015,xu2021}.
	
	From an electronic structure perspective, the challenge of evaluating strongly correlated systems lies in representing the multiconfigurational wavefunction \cite{zhou2022}.
	Addressing this problem has spurred the advent of several computational strategies, each with distinct advantages and limitations. Within the multiconfiguration self-consistent field (MCSCF) framework, complete active space configuration interaction (CASCI) \cite{levine2021} provides a foundation by performing configuration interaction within a predefined active space. Building upon this foundation, the complete active space self-consistent-field (CASSCF) method enhances CASCI by simultaneously optimizing both the CI coefficients and orbital parameters, thereby more accurately capturing static correlation effects——the near-degeneracy effects arising from multiple electronic configurations of comparable energy \cite{olsen2011,wallace2014}. 
	
	However, both CASCI and CASSCF often provide insufficient accuracy due to their incomplete treatment of dynamic correlation effects \cite{olsen2011,pathak2017}——the instantaneous electron-electron repulsion effects \cite{benavides-riveros2017}.
	To remedy this deficiency, researchers have developed various post-SCF approaches that build upon MCSCF reference wavefunctions \cite{helgaker2000}. Complete active space second-order perturbation theory (CASPT2) \cite{roos1996,battaglia2023} represents one of the most successful examples, systematically recovering dynamic correlation, though it suffers from computational scaling limitations.
	
	To address these computational limitations, multiconfiguration pair-density functional theory (MC-PDFT) \cite{limanni2014,ghosh2015} offers a promising alternative by combining CASSCF-derived density matrices with an on-top density functional that capture dynamic correlation through opposite-spin electron pair probabilities at identical spatial coordinates. This hybrid approach greatly improves computational scaling relative to traditional post-SCF methods \cite{zhou2022,gagliardi2017}. 
	
	However, MC-PDFT's efficiency remains fundamentally constrained by its underlying CASSCF calculation. The exponential scaling of CASSCF with active space size renders calculations for large chemical systems computationally prohibitive, even with state-of-the-art hardware and advanced approximation schemes \cite{levine2020}.
	
	The emergence of quantum computing presents a potential solution to these exponential scaling challenges \cite{kandala2017,googleaiquantumandcollaborators2020}. 
	Quantum algorithms naturally exploit superposition and entanglement to encode complex many-body wavefunctions, offering theoretical advantages for strongly correlated electronic systems \cite{tazhigulov2022}.
	Meanwhile, current noisy intermediate-scale quantum (NISQ) devices have motivated the development of hybrid quantum-classical algorithms \cite{preskill2018}, most notably the variational quantum eigensolver (VQE) \cite{peruzzo2014} and its variants. 
	These include ADAPT-VQE \cite{grimsley2019,tang2021}, which dynamically constructs quantum circuits through iterative operator selection; Cluster-VQE \cite{zhang2022d}, which reduces qubit requirements via system decomposition; DMET-ESVQE \cite{li2022}, which achieves the combined advantages of both circuit optimization and qubit reduction through a distinct embedding-based approach; and also ansatz design to maintain physical symmetries including quantum-number-preserving \cite{anselmetti2021}. Alternative frameworks such as quantum algorithms for density functional theory (DFT) have also been proposed \cite{senjean2023}.
	
	For large systems, spatial decomposition approaches offer a practical solution by partitioning the system into fragments amenable to quantum hardware. Recent methods combine fragment molecular orbital techniques with quantum embedding \cite{hardikar2024} and utilize localized active space strategies to reduce quantum resource requirements through systematic treatment of inter-fragment correlations \cite{otten2022,mitra2024}.
	 
	Despite these advances, extending VQE-based methods to multiconfigurational systems remains problematic \cite{sugisaki2022a}. 
	Existing approaches such as UCCGSD, and k-UpCCGSD struggle with the simultaneous treatment of static and dynamic correlation \cite{wecker2015,lee2019}, often requiring prohibitively deep quantum circuits that exceed current NISQ capabilities. 
	Moreover, the qubit counts on near-term devices typically limit such ansatz to compact active spaces, and dynamic correlation contributions from orbitals outside the active space remain missing. Besides, the inherent noise in NISQ devices raises fundamental questions about computational reliability \cite{dasgupta2024,lolur2023} for practical applications, necessitating error analysis in specific calculation.
	Therefore, the fundamental challenge lies in designing quantum algorithms that can efficiently capture multiconfigurational character while remaining compatible with near-term quantum hardware constraints.
	
	An alternative paradigm leverages quantum devices to compute reduced density matrices (RDMs), delegating correlation energy recovery to classical post-processing. Notable implementations range from coupling self-consistent CASSCF with hardware-efficient sampling to mitigate measurement noise in small active spaces \cite{tilly2021}, to integrating VQE with adiabatic connection theory for systematic dynamic correlation recovery from orbitals outside the chosen active space \cite{matousek2024}. 
	Furthermore, Boyn et al. developed a hybrid framework in which a quantum solution of the anti-Hermitian contracted Schrödinger equation directly yields N-representable 2-RDMs, which are then combined with multiconfiguration pair-density functional theory for classical correlation-energy evaluation \cite{boyn2021}.
	
	Collectively, the above RDM-based approaches support the viability of separating quantum wavefunction preparation from classical correlation recovery. Building upon this philosophy, we introduce a quantum-classical hybrid framework that integrates VQE with multiconfiguration pair-density functional theory (MC-PDFT), referred to as VQE-PDFT. This framework (described first below) then serves as the basis for two additional developments targeting realistic biological modeling and practical hardware execution.
	
	First, at the methodological level, we employ VQE strictly as a CASCI solver to capture static correlation, while recovering additional dynamic correlation beyond the CASCI description via a classical MC-PDFT on-top functional. In this framework, the total energy is evaluated from a pair-density functional of CASCI-level RDMs rather than directly from the Hamiltonian expectation value, allowing the quantum computation in VQE-PDFT to be reduced to compact active spaces focused on static correlation. By contrast, bare-$\langle \mathrm{H} \rangle$ VQE approaches may access a comparable level of dynamic correlation by enlarging the active space mapped to qubits and/or employing highly expressive ansatz (e.g., UCCGSD and k-UpCCGSD \cite{wecker2015,lee2019}), leading to increased qubit requirements and circuit depth.
	
	Second, we embed VQE-PDFT within a multiscale quantum mechanics/molecular mechanics (QM/MM) workflow to study electron transfer in a cryptochrome, computing Marcus parameters and transfer rates from ensembles of protein conformations in a complex biological environment.
	
	Third, we develop shallow, symmetry-preserving hardware-efficient ansatz tailored to open- and closed-shell tryptophan active spaces of the cryptochrome electron-transfer center, further reducing qubit count and circuit depth to support an initial hardware demonstration.
	
	In this study, we first validated VQE-PDFT on the CT7/04 Charge-Transfer benchmark dataset, confirming accuracy comparable to conventional MC-PDFT. We then presented its QM/MM application to electron transfer in the European robin cryptochrome protein (ErCRY4), where the transfer rates from noiseless simulations agreed well with ultrafast spectroscopy measurements \cite{timmer2023}. Finally, as a proof-of-principle hardware validation on a simplified active-space model, a single conformation experiment on a 13-qubit superconducting device illustrated feasibility on current NISQ hardware within well-defined resource limits.
	
	Throughout this work, unless explicitly stated otherwise, most of the quantum circuit calculations were carried out on noiseless classical simulators in limited active spaces, with quantum hardware used only for a single proof-of-principle validation.
	
	\section{Results}\label{results}
	
	\subsection{The VQE integrated MC-PDFT}
	
	Our quantum-classical hybrid approach VQE-PDFT replaces the computationally expensive CASSCF optimization in MC-PDFT with a VQE solver for the CASCI active-space Hamiltonian. 
	In all calculations reported here, this CASCI Hamiltonian is constructed using Hartree–Fock canonical molecular orbitals for the chosen active spaces, without further orbital optimization  included. Consequently, the VQE state should be viewed as a CASCI-level wavefunction rather than a fully self-consistent CASSCF state.
	
\begin{figure}[htb]
		\centering
		\includegraphics[width=0.45\textwidth]{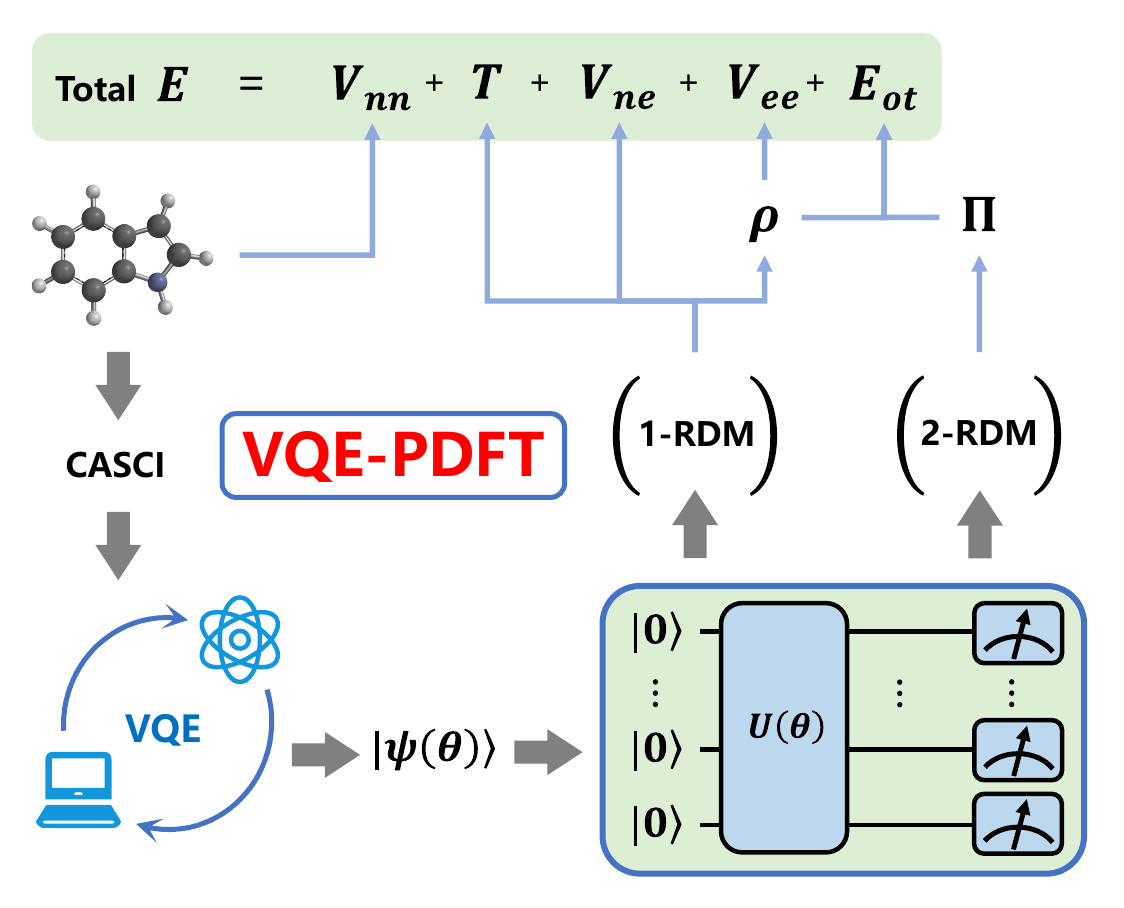}
		\caption{\textbf{Workflow of VQE-PDFT.} Quantum computing is used as a solver for the CASCI active-space Hamiltonian and for the subsequent evaluation of 1-RDM and 2-RDM elements, where the “CASCI” denotes the definition of the CASCI active-space Hamiltonian and the associated one- and two-electron integrals. The VQE loop and the quantum circuit $U(\theta)$ below it represent the numerical solution of this CASCI Hamiltonian, yielding the state $|\psi(\theta)\rangle$ from which the 1- and 2-RDMs are sampled and passed to further calculate the PDFT energy. Here, “CASCI” refers to the level of theory of the active-space Hamiltonian, while the particular VQE ansatz (UCCSD/ROUCCSD/HEA) specifies how the CASCI wavefunction is parameterized on the quantum circuit.
		}
		\label{fig:vqe-pdft}
	\end{figure}

	The modified workflow proceeds as follows: VQE optimizes a parameterized quantum circuit ansatz (UCCSD for the CT7/04 benchmark, and ROUCCSD or the empirical HEA circuits for the ErCRY4 application) to approximate the CASCI ground state in the chosen active space, from which one-particle and two-particle reduced density matrices (1-RDM and 2-RDM) are extracted by measuring the corresponding matrix element expectation values on the quantum circuit as shown in Fig. \ref{fig:vqe-pdft}. These reduced density matrices are then utilized to compute the total energy following the MC-PDFT formalism.
	
	The 1-RDM provides the kinetic energy $T$, nuclear-electron interaction $V_{ne}$, total electron density $\rho$, and classical Coulomb repulsion $V_{ee}(\rho)$. The 2-RDM yields the on-top pair density $\Pi$, which combines with $\rho$ to determine the on-top density functional energy $E_{ot}(\rho,\Pi)$ that captures exchange-correlation effects. Finally, together with the nuclear-nuclear repulsion $V_{nn}$, the total energy follows the standard MC-PDFT expression:
	\begin{equation}\label{e_tot}
	E=T+V_{ne}+V_{nn}+V_{ee}(\rho)+E_{ot}(\rho,\Pi).
	\end{equation}
	
	This hybrid strategy is more compatible with NISQ constraints because the quantum computation is confined to a CASCI active-space solver that captures static correlation, while dynamic correlation is recovered classically through the on-top functional \cite{limanni2014}. It therefore avoids treating both correlation regimes variationally on the quantum circuit, which would generally require a highly expressive yet prohibitively deep ansatz (e.g., UCCGSD \cite{wecker2015}). Moreover, capturing large part of the dynamic correlation beyond a compact active space would require an expanded orbital space, further increasing the qubit counts.
	
	\subsubsection*{Benchmarks on dissociation energy}
	
	To validate our VQE-PDFT approach, we evaluated its performance on the CT7/04 dataset \cite{zhao2005} with noiseless simulated quantum circuit. To ensure computational feasibility, we employed a reduced active space up to (10e,10o), requiring up to 20 qubits (see Table \ref{tab:ct7}). Detailed quantum resource information for each dimer and monomer is summarized in Supplementary Table S1. 
	
	This dataset comprises 7 dimers that exhibit significant charge transfer effects upon dissociation. These systems are notoriously difficult to calculate accurately due to two key factors: their inherent multiconfigurational character and the dramatic electronic reorganization that occurs during dissociation \cite{vitillo2022,lischka2018}. Such characteristics make these dimers ideal test cases for evaluating multiconfigurational methods in further applications \cite{ghosh2015}.
	
	\begin{table*}[ht]
		\centering
		\begin{tabular}{cccccc}
			\toprule
			Dimers & $N_q$ (dimer) & VQE & VQE-PDFT & MC-PDFT & W1-reference \\
			\midrule
			\( NH_3 \cdots FCl \)     & 20 &  5.38 & 11.56 & 12.42 & 10.62 \\
			\( NH_3 \cdots Cl_2 \)    & 20 &  1.57 & 5.09  & 4.55  & 4.88  \\
			\( NH_3 \cdots F_2 \)     & 20 & -0.08 & 1.88  & 1.11  & 1.81  \\
			\( HCN \cdots FCl \)      & 16 &  2.19 & 3.13  & 3.67  & 4.86  \\
			\( H_2O \cdots FCl \)     & 20 & 10.01 & 3.30  & 4.38  & 5.36  \\
			\( C_2H_2 \cdots FCl \)   & 20 &  4.13 & 3.07  & 4.01  & 3.81  \\
			\( C_2H_4 \cdots F_2 \)   & 20 & -1.14 & 0.85  & 0.33  & 1.06  \\
			\midrule
			\multicolumn{2}{c}{MUE} & 2.896 & 0.853 & 0.847 & -- \\
			\bottomrule
		\end{tabular} 
		\captionsetup{width=\linewidth}
		\caption{\textbf{Dissociation energy of Charge-Transfer dimers dataset CT7/04.} Energies are given in kcal/mol. The geometries of dimers and their reference values were fetched from Ref. \cite{zhao2005, zhao2005a}. Here, the column labeled `VQE' corresponds to the CASCI energy $\langle \psi(\theta) | \hat{H}_\text{CASCI} | \psi(\theta) \rangle$ evaluated from the same VQE-optimized active-space wavefunction that is used to construct the RDMs in VQE-PDFT. 
		All `VQE' and `VQE-PDFT' results were obtained from noiseless classical simulations of UCCSD ansatz in limited active spaces (see Supplementary Table S1). The column “$N_q$ (dimer)” reports the number of qubits used for each dimer calculation (20 qubits for six systems and 16 qubits for one system). The detailed active space specifications and quantum resource summary (for both dimers and monomers) are provided in the Supplementary Table S1. 
		The results of MC-PDFT were obtained from Ghosh et al.'s research \cite{ghosh2015}. The jul-cc-pVTZ \cite{papajak2011} basis set was used in all calculations. The tPBE functional was adopted in VQE-PDFT, which aligned with Ref. \cite{ghosh2015}. The MUE stands for the mean unsigned error. 
		More explicit absolute energies of every systems are listed in Supplementary Table S3. 
		}
		\label{tab:ct7}
	\end{table*}
	
	Table \ref{tab:ct7} summarizes the CT7/04 dissociation energies obtained with CASCI-level VQE (`VQE'), VQE-PDFT, and classical MC-PDFT, together with high-level Weizmann-1 theory reference values (`W1-reference') \cite{zhao2005}. Utilizing the Unitary Coupled-Cluster Singles and Doubles (UCCSD) ansatz, our VQE-PDFT method achieved a mean unsigned error (MUE) of 0.853 kcal/mol relative to W1 theory, closely matching the classical MC-PDFT's result of 0.847 kcal/mol \cite{ghosh2015}. 
	
	We emphasize that this close agreement between VQE-PDFT and classical MC-PDFT for CT7/04 was achieved within the specific limited active spaces and numerical settings adopted here (Supplementary Table S1), and different active-space choices or numerical approximations could modify the quantitative agreement even though both protocols share the same on-top functional.
	
	Notably, for two dimers ($NH_3 \cdots F_2$ and $C_2H_4 \cdots F_2$), the CASCI-level VQE (column `VQE' in Table \ref{tab:ct7}) yielded negative dissociation energies. This indicated that, within these active spaces, the bare CASCI-level energies could be insufficiently correlated for these charge-transfer interactions. When the identical VQE-derived 1- and 2-RDMs were instead combined with the MC-PDFT functional (VQE-PDFT column), the dissociation energies restored the correct sign and moved closer to the reference values, and the MUE was reduced to 0.853 kcal/mol. This comparison implied that the improvement of VQE-PDFT over CASCI-level VQE arose primarily from the functional treatment of dynamic correlation, instead of a change in the underlying VQE state or additional quantum resources.
	
	This interpretation is consistent with the analysis of Ghosh et al. for CT7/04, where classical CASSCF (an orbital-optimizing version of CASCI) was found to have the largest MUE (3.92 kcal/mol) among the methods considered and yielded negative dissociation energies on $C_2H_2 \cdots FCl$ and $C_2H_4 \cdots F_2$ dimers in CT7/04. Ghosh et al. attributed this to the absence of an explicit approximation to the full dynamic correlation energy \cite{ghosh2015}. Accordingly, the CASCI-level energies in the `VQE' column may mainly reflect the known limitations of the CASCI framework with respect to dynamic correlation, rather than an intrinsic deficiency of the VQE algorithm itself. Conversely, the improved VQE–PDFT dissociation energies result from supplementing this missing dynamic correlation through the MC-PDFT functional applied to the same VQE-derived RDMs (even though the resulting energy is no longer a strict variational $\langle \Psi | \hat{H} | \Psi \rangle$ for that state).
	
	From this perspective, the `VQE' column in Table \ref{tab:ct7} may be viewed as the CASCI-level analogue of the classical CASSCF reference step in conventional MC-PDFT, whereas the `VQE-PDFT' and `MC-PDFT' columns correspond to applying the same on-top pair-density functional to quantum- and classically obtained multiconfigurational references, respectively.
	
	To better understand the error patterns, we further analyzed the deviations besides the MUE metric. The root-mean-square errors (RMSE) relative to W1 reference were 1.129 kcal/mol for VQE-PDFT and 0.985 kcal/mol for MC-PDFT, showing that VQE-PDFT exhibited a slightly broader error distribution yet averaged out in the MUE metric. Additionally, inspection of the absolute total energies (Supplementary Table S3) revealed a method-dependent systematic shift of VQE-PDFT compared to MC-PDFT results. However, since the dissociation energy is defined as an energy difference ($E_{\mathrm{dimer}} - \sum E_{\mathrm{monomer}}$), these systematic shifts partially canceled out, leaving residual errors that were comparable to the classical MC-PDFT reference. This suggested that the agreement in dissociation energies reflects a combination of similar dynamic correlation treatment via the tPBE functional and the beneficial cancellation of systematic offset.
	
	In summary, this benchmark suggests that VQE-PDFT may effectively describe the electronic correlation in these charge-transfer systems. This indicates its potential for handling similar complex interactions found in bio-molecules, where multiconfigurational character is ubiquitous. Moreover, the observation that method-dependent shifts partially canceled in energy differences provides the rationale for our subsequent application to the electron transfer process in ErCRY4 protein. Since the key parameters (reorganization energy and driving force in Marcus theory) are also defined as differences between single-point energies, we anticipate a similar degree of error cancellation, enabling reliable predictions of electron transfer rates even within simplified active-space models.
	
	\subsection{Quantum-classical hybrid framework for biological electron transfer}
	\begin{figure*}[t]
		\centering
		\includegraphics[width=1\textwidth]{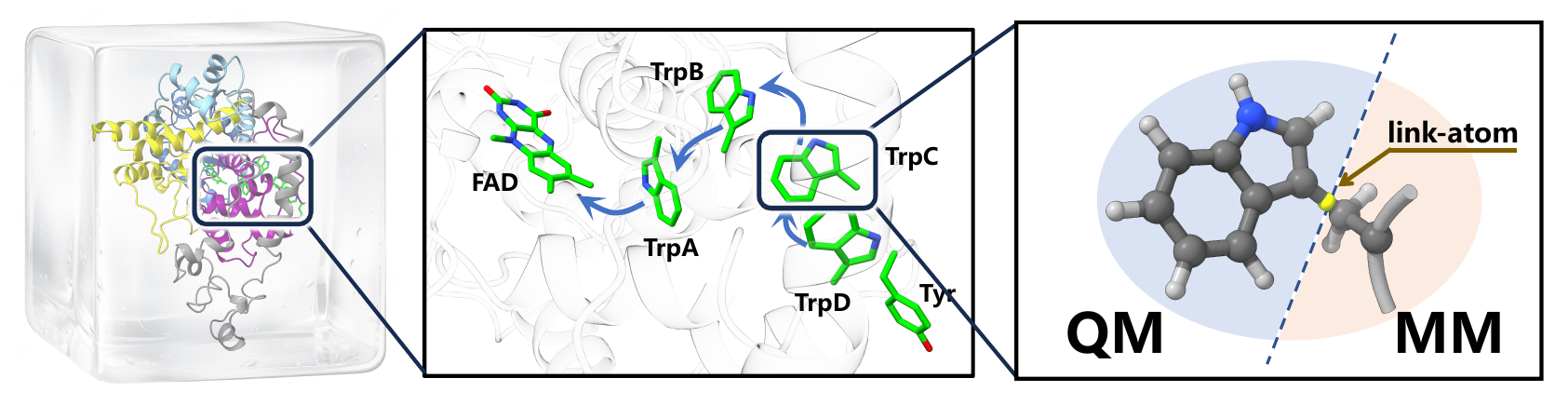}
		\caption{\textbf{Multiscale calculation framework of ErCRY4 protein.} The quantum mechanical (QM) region is set to be two indole rings on the adjacent tryptophan residues (TrpB and TrpC), which are treated in separate QM calculations, while the rest of the systems are considered in molecular mechanical (MM) level. The standard electrostatic embedding and link-atom methodology are adopted in evaluating interactions between the QM and MM regions. The figures were drawn with the help of ChimeraX \cite{meng2023}.}
		\label{fig:qmmm}
	\end{figure*}
	
	Building upon the validated accuracy of our VQE-PDFT method, we developed a comprehensive framework to apply quantum-classical hybrid calculations to a representative biological system, focusing on the electron transfer process in ErCRY4. 
	Because in Marcus Theory, the reorganization energy and driving force are constructed from differences of closely related single-point energies, therefore the previously observed partial cancellation of method-dependent shifts in energy differences now provides a direct motivation for the following transfer rate calculations.
	
	Overall, this framework addresses two key challenges: adapting a multiscale QM/MM architecture, based on standard electrostatic embedding and link-atom techniques, for complex biological environments; and designing hardware-efficient quantum circuits optimized for NISQ constraints in electron transfer calculations.
	
	Regarding the multiscale coupling, the QM/MM machinery itself follows well-established practice. In our workflow, the quantum circuit is used specifically as a CASCI solver for the multiconfigurational QM region, replacing the classical CASSCF step within an otherwise MC-PDFT/QM/MM treatment. Because the present ErCRY4 proof-of-concept employs deliberately compact active spaces that also classically tractable, we do not claim a demonstrated quantum speed-up for this application. Rather, the motivation is that, as qubit numbers and gate fidelities improve, a quantum CASCI solver could help alleviate the main bottleneck of classical multiconfigurational solvers, and thereby enable larger active spaces in the QM region within the same multiscale framework. This long-term perspective is consistent with quantum resource-estimate studies of other strongly correlated bioinorganic centers, such as the nitrogenase FeMo cofactor, where Fe--S clusters pose challenges for conventional multiconfigurational treatments and have been proposed as natural targets for future quantum simulations.\cite{reiher2017, li2019}
	
	\subsubsection{QM/MM multiscale architecture}
	
	Inheriting from the hybrid quantum computing pipeline previously developed by Li et al. \cite{li2024a}, the multiscale architecture employs a QM/MM partitioning scheme where the quantum region, containing the multiconfigurational active sites, is treated by our VQE-PDFT method, while the surrounding protein environment and solvent are described using classical molecular mechanics force fields, as illustrated in Fig. \ref{fig:qmmm}.
	
	Extending this architecture for the subsequent study, the standard electrostatic embedding and link-atom methodology are implemented to handle the interface between QM and MM regions. Especially, it involves capping the covalent `C-C' bond cut at the QM/MM boundary, ensuring proper treatment of boundary effects \cite{senn2009}. 
	
	To provide a proof-of-concept application of this multiscale framework, we selected the electron transfer process between adjacent tryptophan residues (TrpB and TrpC in Fig. \ref{fig:qmmm}) in the European robin cryptochrome protein (ErCRY4) as our target study.
	This system presents an ideal test case due to its well-characterized experimental properties \cite{xu2021,timmer2023} and the multiconfigurational nature. 
	The electron transfer process involves photo-excitation of the FAD cofactor followed by sequential electron transfer among adjacent tryptophan residues, during which tryptophan residues alternate between neutral and cationic states. 
	
	In this QM/MM setup, the QM region is represented by two separated fragments corresponding to the indole rings of TrpB and TrpC. Each single-point QM calculation is carried out on one fragment (TrpB or TrpC), while the remainder of the amino-acid side chain is treated at the MM level. For each QM fragment, the VQE-PDFT calculations use a compact CASCI active space consisting of three frontier orbitals localized on the tryptophan indole ring (Supplementary Figure S1), with 4 or 3 active electrons for the neutral and cationic states, respectively. The corresponding CASCI Hamiltonian and active orbitals are generated by the CASCI module of PySCF \cite{sun2020} as interfaced in TenCirChem \cite{li2023c}, starting from Hartree–Fock canonical orbitals and using the default active-space construction based on the specified $(N_\mathrm{e},N_\mathrm{o})$ pattern.
	
	However, the cationic states exhibit open-shell electronic character with unpaired electrons, making the previously used standard UCCSD ansatz inadequate, as it assumes paired $\alpha$ and $\beta$ electrons in each spatial orbital. 
	This limitation, combined with NISQ device constraints, necessitates the development of specialized quantum circuits.
	
	\subsection{Shallow-depth empirical ansatz design}
	Conventional UCCSD ansatz performs well on evaluating closed-shell systems, and can be extended to restricted open-shell UCCSD (ROUCCSD) by explicitly considering single-electron occupations and the corresponding excitations to accommodate the unpaired electron scenarios in cationic tryptophan. However, ROUCCSD's prohibitively deep circuit structure and extensive non-local gate operations render it impractical for NISQ device implementation. To address this limitation, we developed shallow depth empirical hardware-efficient ansatz (HEA) circuits specifically tailored for the electronic structure and symmetry of the tryptophan system.
	 
	\begin{figure}[htb]
		\centering
		\includegraphics[width=0.4\textwidth]{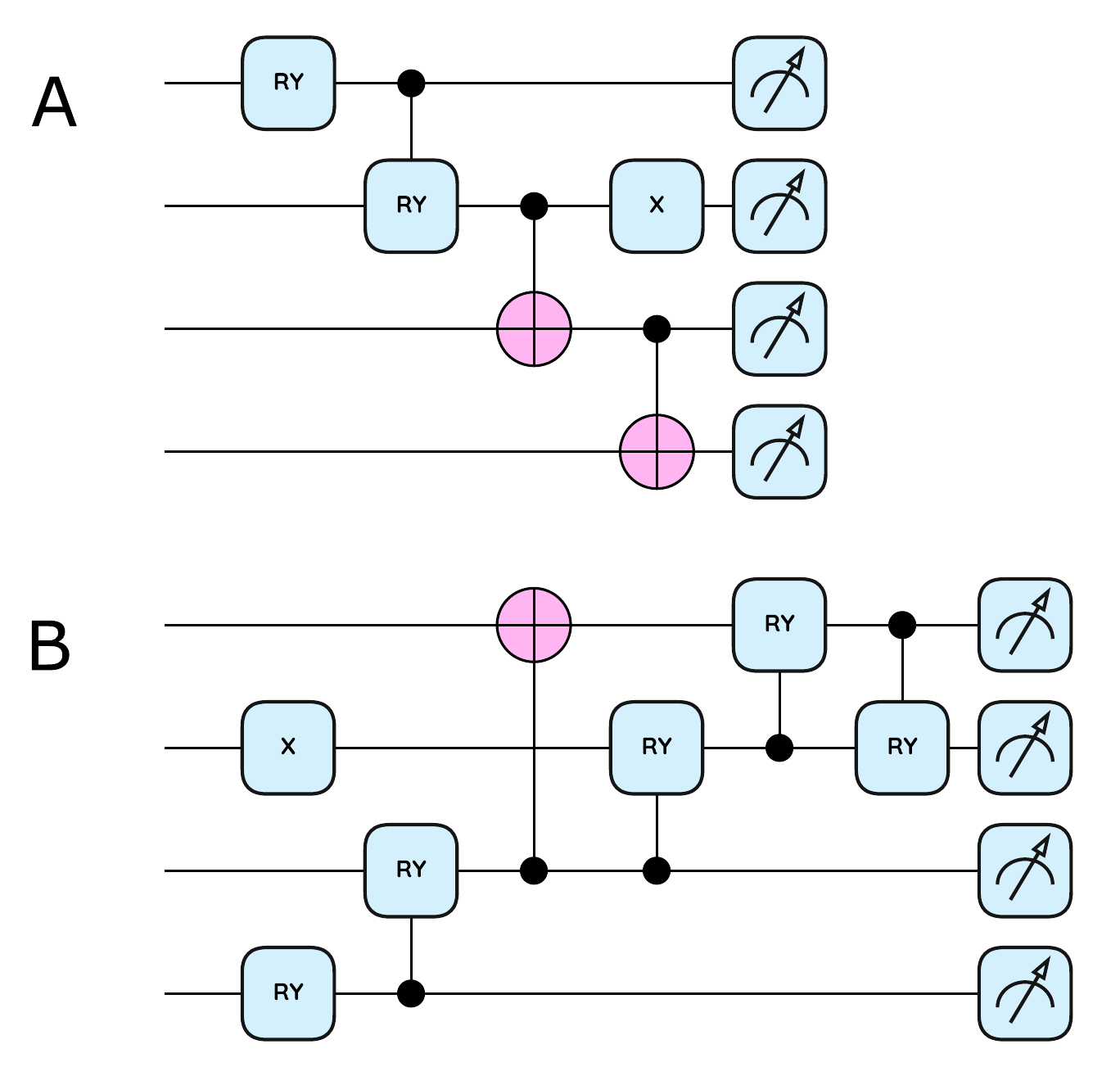}
		\caption{\textbf{The shallow-depth empirical ansatzes for electron transfer.} \textbf{A} Closed-shell HEA circuit (CHEA) with circuit depth of 4. \textbf{B} Open-shell HEA circuit (OHEA) with circuit depth of 6. These two circuits are dramatically shallower than ROUCCSD with circuit depth of 35. The circuit diagrams were drawn with the PennyLane library \cite{bergholm2022}.
		}
		\label{fig:hea}
	\end{figure}
	Our approach employs distinct quantum circuits for the two electronic states encountered in the transfer process: a closed-shell HEA (CHEA, Fig. \ref{fig:hea} A) for neutral tryptophan and an open-shell HEA (OHEA, Fig. \ref{fig:hea} B) for cationic tryptophan. Both circuits are empirically designed to maintain the particle number conservation for each spin (i.e., within the target $N_{\alpha}/N_{\beta}$ symmetry sector after tapering), while achieving minimal circuit depth compared to the ROUCCSD ansatz. The designing methodology is detailed in Section \ref{EHEA method}.
	
	To validate the accuracy of the empirical HEA circuits, we randomly extracted a protein conformation from ErCRY4 molecular dynamics simulations reported in Ref. \cite{xu2021}. We then applied our multiscale framework by defining two adjacent tryptophan residues as separated QM regions for computational analysis.
	We employed a systematic approach that generated eight single-point energy calculations for two tryptophan residues, encompassing four open-shell cationic states and four closed-shell neutral states. These states are typically encountered in electron transfer processes, and the specific procedure of constructing these states follows the four-point scheme \cite{lopez-estrada2018} (see also section \ref{fourpoint}). 
	
	For this validation study, we performed calculations using three different methods: the empirical HEA, ROUCCSD, and full configuration interaction (FCI), each applied to the active space within the CASCI framework. The comparative results are presented in Table \ref{tab:hea}.
	
	\begin{table}[ht]
		\centering
		\begin{tabular}{ccccc}
			\toprule
			\text{ } & \text{$E_{FCI}$} & \text{$E_{ROUCCSD}$} & \text{$E_{HEA}$} & \text{$\Delta E_{HEA-FCI}$} \\
			\midrule
			1   & -2.1968    & -2.1968    & -2.1947    & 0.0021    \\
			2   & -2.5535    & -2.5535    & -2.5500    & 0.0035    \\
			3   & -2.1902    & -2.1902    & -2.1899    & 0.0004    \\
			4   & -2.5424    & -2.5424    & -2.5391    & 0.0033    \\
			5   & -2.1912    & -2.1912    & -2.1899    & 0.0013    \\
			6   & -2.5621    & -2.5621    & -2.5584    & 0.0036    \\
			7   & -2.1963    & -2.1963    & -2.1958    & 0.0005    \\
			8   & -2.5310    & -2.5310    & -2.5279    & 0.0032    \\
			\bottomrule
		\end{tabular}
		\caption{\textbf{Energy calculations on the active space of tryptophan.} The 1st, 3rd, 5th, and 7th single-point calculations are of open-shell with (3e,3o) active space, while the others are of closed-shell with (4e,3o) active space. The 6-31G basis set was used in all calculations. All energies are given in Hartree.}
		\label{tab:hea}
	\end{table}
	
	The results demonstrate that the proposed empirical HEA circuits achieve accuracy of energy differences typically ranging from 10\textsuperscript{-4} to 10\textsuperscript{-3} Hartree with respect to the FCI's results. Moreover, ROUCCSD energies match FCI values within excellent precision (at least $<10^{-4}$ Hartree). We note that this near identity between ROUCCSD and FCI may arise from the compact (3e,3o) and (4e,3o) active spaces adopted in this validation, and should not be interpreted as a general property of ROUCCSD for larger multiconfigurational active spaces, where it remains an approximate high-level reference to FCI.
	
	Also in Table \ref{tab:hea}, one may observe the slightly smaller HEA-FCI deviations for the open-shell cases (rows 1, 3, 5, and 7), compared to closed-shell's results. This is consistent with our deliberate circuit design: to accommodate the more complex open-shell configurations, OHEA is constructed with a modestly larger depth and more variational parameters than CHEA (depth 6 vs 4, and 5 vs 2 parameters; Fig. \ref{fig:hea}).
	At the same time, the empirical HEA circuits are intentionally constrained to a fixed shallow depth and a small number of variational parameters to prioritize NISQ feasibility. Therefore, the residual HEA-FCI discrepancies in Table \ref{tab:hea} mainly reflect this resource-oriented ansatz choice rather than an intrinsic limitation of the general HEA framework.
	
	Overall, this validation confirms that our HEA circuits maintain sufficient expressibility for the active space of current electron transfer system while offering notable quantum resource savings. 
	
	\subsection{Evaluate electron transfer in the ErCRY4}
	
	\subsubsection{Marcus theory}
	
	Having demonstrated the accuracy of the empirical HEA circuits, we applied them to compute electron transfer rates in the ErCRY4 protein system. Electron transfer kinetics in biological systems may be depicted by the Marcus theory, with the transfer rate expressed as:
	\begin{equation}\label{marcustheory}
	k_{ET} = \frac{2\pi}{\hbar}\frac{\left\langle |H_{DA}|^2\right\rangle}{\sqrt{4\pi \lambda k_B T}}\exp\left[-\frac{(\Delta G^o + \lambda)^2}{4\lambda k_B T}\right],
	\end{equation}
	which incorporates three key parameters: $\left\langle |H_{DA}|\right\rangle$, $\lambda$, and $\Delta G^o$. The electronic coupling $\left\langle |H_{DA}|\right\rangle$ represents the interaction between two tryptophan residues, reflecting the orbital overlap and electronic communication pathways. Then the reorganization energy $\lambda$ quantifies the energetic cost of nuclear rearrangement accompanying electron transfer. Lastly, the driving force $\Delta G^o$ corresponds to the thermodynamic bias for the transfer process, determined by the energy difference between reactant and product states.
	
	Our VQE-PDFT framework within the multiscale QM/MM architecture enables direct evaluation of the reorganization energy and driving force.
	
	\subsubsection{Four-point scheme for $\lambda$ and $\Delta G^o$}\label{fourpoint}
	
	With the Marcus theory framework outlined above, we now detail our computational approach. We focused specifically on the electron transfer between tryptophan residues TrpB (W372) and TrpC (W318) in the ErCRY4 protein. The initial state corresponded to cationic TrpB and neutral TrpC, while the final state involved neutral TrpB and cationic TrpC following electron transfer.
	
	\begin{figure}[htb]
		\centering
		\includegraphics[width=0.45\textwidth]{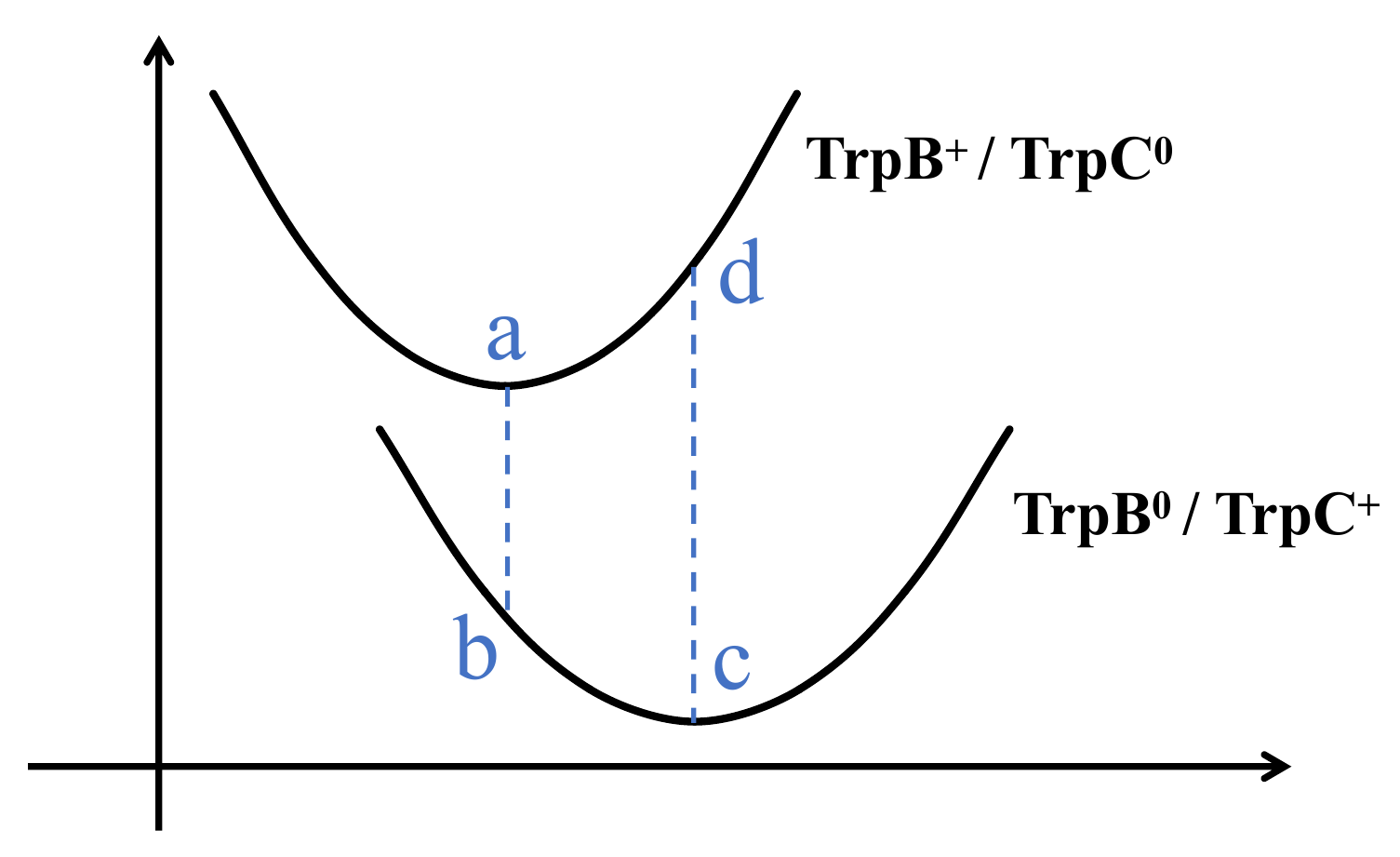}
		\caption{\textbf{The four-point scheme for evaluating reorganization energy and driving force.} a: TrpB\textsuperscript{+}/TrpC\textsuperscript{0} state on the initial geometry; b: TrpB\textsuperscript{0}/TrpC\textsuperscript{+} state on the initial geometry; c: TrpB\textsuperscript{0}/TrpC\textsuperscript{+} state on the final geometry; d: TrpB\textsuperscript{+}/TrpC\textsuperscript{0} state on the final geometry.}
		\label{fig:fourpoint}
	\end{figure}
	
	To quantify both the reorganization energy $\lambda$ and driving force $\Delta G^o$, we employed the four-point scheme \cite{lopez-estrada2018}, a well-established thermodynamic approach that captures the essential physics of Marcus theory. This method is based on the fundamental assumption that the system responds instantaneously to changes in electronic charge distribution \cite{marcus1985}. This approximation, although neglects potential non-Markovian effects arising from slower environmental responses \cite{firmino2016}, enables us to bypass costly dynamical simulations and focus directly on energetic differences between key electronic configurations.
	
	As illustrated in Fig. \ref{fig:fourpoint}, the four-point scheme evaluates single-point energies at four distinct state-geometry combinations: each electronic state (initial and final) calculated at both its own optimized nuclear conformation and at the geometry optimized for the other state. This approach captures the energetic penalty arising from the mismatch between optimal nuclear arrangements for different charge distributions. Thereafter, the reorganization energy and the driving force of a given protein conformation can be extracted through:
	\begin{align}
		\Delta G^o &= \left|E_{i}^{i} - E_{f}^{f}\right|, \label{eq:deltag}\\
		\lambda &= \left| E_{f}^{i} - E_{i}^{i} \right| + \left| E_{i}^{f} - E_{f}^{f} \right|, \label{eq:lambda}
	\end{align}
	where $E_{state}^{geometry}$ denotes the energy of a given electronic state calculated at a specific nuclear geometry, while ``$i$" and ``$f$" stands for ``initial" and ``final" respectively.
	
	\begin{table}[ht]
		\centering
		\begin{tabular}{|c|c|c|c|c|}
			\hline
			Ansatz        & $\lambda$    & $\Delta G^o$ & $\left\langle |H_{DA}|\right\rangle$ & $\left\langle |H_{DA}|^2\right\rangle$\\ \hline
			ROUCCSD       & 0.5701       & 0.0689       & \multirow{2}{*}{6.352E-3}           & \multirow{2}{*}{1.1431E-4}            \\ \cline{1-3}
			HEA & 0.4356       & 0.0724       &                                      &                                       \\ \hline
		\end{tabular}
		\caption{\textbf{The averaged $\lambda$, $\Delta G^o$, and $\left\langle |H_{DA}|\right\rangle$.} The reorganization energy and driving force were evaluated by four-point scheme, utilizing ROUCCSD as well as the empirical HEA, while the electronic coupling were determined by direct coupling scheme. The detailed results for individual configurations are provided in the Supporting Information. $\lambda$, $\Delta G^o$, and $\left\langle |H_{DA}|\right\rangle$ are given in eV, while $\left\langle |H_{DA}|^2 \right\rangle$ is in (eV)\textsuperscript{2}.The 6-31G basis set and the tPBE functional were used in all calculations.}
		\label{tab:param1}
	\end{table}
	
	To capture the dynamic behavior of the protein in biological environments, we sampled 20 distinct protein conformations from molecular dynamics simulations of ErCRY4 and applied the four-point scheme to each conformation using our VQE-PDFT/MM framework. We calculated $\lambda$ and $\Delta G^o$ using both ROUCCSD and the proposed empirical HEA circuits, obtaining averaged value of $\lambda = 0.5701$ eV and $\Delta G^o = 0.0689$ eV for ROUCCSD, and $\lambda = 0.4356$ eV and $\Delta G^o = 0.0724$ eV for HEA, as listed in Table \ref{tab:param1} and detailed in Supplementary Information. The good agreement between HEA and ROUCCSD results, considering the great reduction in circuit depth in HEA, validates the feasibility of the empirical HEA for the current biological application, achieving a balance between accuracy and quantum resource efficiency, although with a compact active space.
	
	\subsubsection{Electronic coupling}
	
	With the reorganization energy and driving force determined, we now evaluate the electronic coupling $\left\langle |H_{DA}|\right\rangle$. 
	
	We computed the $\left\langle |H_{DA}|\right\rangle$ through transfer integral calculations between the TrpB and TrpC indole rings. The approach employed Boys localization \cite{subotnik2008} to transform canonical molecular orbitals into spatially localized orbitals centered on individual tryptophan sites, enabling clear identification of the relevant frontier orbitals for each fragment. This localization procedure was essential for defining well-separated donor and acceptor orbitals in weakly coupled systems where standard canonical orbitals are typically delocalized across multiple sites.
	
	The electronic coupling is extracted using the direct coupling (DC) scheme \cite{hsu2009}, which accounts for orbital non-orthogonality through:
	\begin{equation}
		H_{DA} = \left[T_{DA} - (e_D + e_A)S_{DA}/2\right]/(1-S_{DA}^2),
	\end{equation}
	where $T_{DA}$ is the Hamiltonian coupling element, $S_{DA}$ is overlap integral, and $e_D, e_A$ are localized orbital energies for donor and acceptor sites, respectively.
	
	Applying this methodology to the same 20 protein configurations, we obtained an averaged electronic coupling of $\left\langle |H_{DA}|\right\rangle = 6.35 \times 10^{-3}$ eV (6.35 meV) that aligned well with the 5 meV reported by Timmer et al. \cite{timmer2023}, also listed in Table \ref{tab:param1}.
	
	\subsubsection{Electron transfer rate}
	
	With all Marcus theory parameters determined from the 20 sampled protein configurations, we proceeded to evaluate the electron transfer rates using Equation (\ref{marcustheory}). 
	
\begin{figure}[htb]
		\centering
		\includegraphics[width=0.45\textwidth]{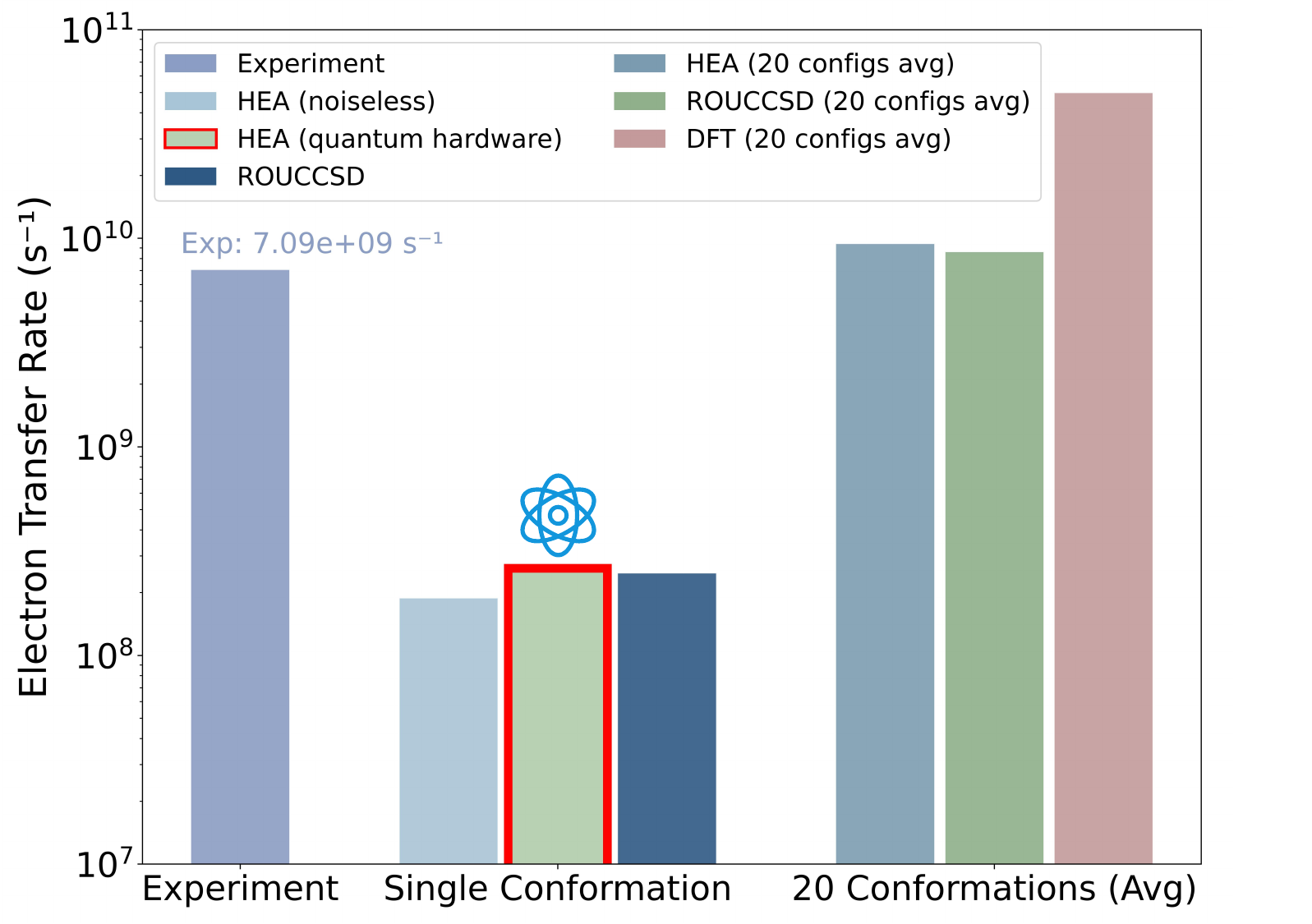}
		\caption{\textbf{Electron transfer rate comparison.} The red marked transfer rate was evaluated on a quantum hardware. The single conformation results show rates one order of magnitude lower than averaged values of 20 conformations, reflecting typical conformational fluctuations in MD simulations where tryptophan positioning variations can alter transfer rates by several orders of magnitude \cite{firmino2016}. The DFT prediction was obtained from Ref. \cite{xu2021}. Explicit parameters calculated for each conformations are listed in Supplementary Information.}
		\label{fig:etrate}
	\end{figure}

	The computed rates were derived by substituting the averaged reorganization energies, driving forces, and electronic couplings into the Marcus expression (\ref{marcustheory}). As a result, our empirical HEA approach yielded $k_{ET} = 0.944 \times 10^{10} s^{-1}$, aligning well with both ROUCCSD calculations ($0.864 \times 10^{10} s^{-1}$) and the experimental ultrafast transient absorption measurements \cite{timmer2023} ($0.709 \times 10^{10} s^{-1}$), as shown in Fig. \ref{fig:etrate}. Notably, these results were approximately one order of magnitude slower than previous DFT predictions \cite{xu2021} ($5.0 \pm 1.8 \times 10^{10} s^{-1}$), as also pointed out in Ref. \cite{timmer2023}.
	
	\subsubsection*{Experimental validation on quantum hardware}
	
	Moreover, to validate our approach on actual NISQ hardware, we executed the empirical HEA circuits on a superconducting quantum device for a randomly selected protein conformation. With VQE's optimization performed on a noiseless simulator and density matrix evaluations on quantum hardware with readout error mitigation \cite{bravyi2021,lolur2023}, we obtained the single-point energies for the four-point scheme, eventually leading to the transfer rate $k_{ET} \approx 2.62 \times 10^8 s^{-1}$, as listed in Table \ref{tab:2071}. 
	
	\begin{table}[ht]
		\centering
		\begin{tabular}{|cc|c|}
			\hline
			\multicolumn{1}{|c|}{\textbf{Index}} & \textbf{System}                 & \textbf{\begin{tabular}[c]{@{}c@{}}Empirical HEA on\\ quantum hardware\end{tabular}} \\ \hline
			\multicolumn{1}{|c|}{1}              & point-a TrpB\textsuperscript{+} & -362.92966                                                                           \\ \hline
			\multicolumn{1}{|c|}{2}              & point-a TrpC\textsuperscript{0} & -363.17866                                                                           \\ \hline
			\multicolumn{1}{|c|}{3}              & point-b TrpB\textsuperscript{0} & -363.20554                                                                           \\ \hline
			\multicolumn{1}{|c|}{4}              & point-b TrpC\textsuperscript{+} & -362.91475                                                                           \\ \hline
			\multicolumn{1}{|c|}{5}              & point-c TrpB\textsuperscript{0} & -363.19954                                                                           \\ \hline
			\multicolumn{1}{|c|}{6}              & point-c TrpC\textsuperscript{+} & -362.91126                                                                           \\ \hline
			\multicolumn{1}{|c|}{7}              & point-d TrpB\textsuperscript{+} & -362.92451                                                                           \\ \hline
			\multicolumn{1}{|c|}{8}              & point-d TrpC\textsuperscript{0} & -363.18130                                                                           \\ \hline
			\hline
			\multicolumn{2}{|c|}{$\lambda$}                                        & 0.47021 $eV$                                                                         \\ \hline
			\multicolumn{2}{|c|}{$\Delta G^o$}                                     & 0.06764 $eV$                                                                         \\ \hline
			\multicolumn{2}{|c|}{$k_{ET}$}                                         & 2.61742$\times 10^8 /s$                                                              \\ \hline
		\end{tabular}
	\caption{\textbf{Single-point energies, Marcus parameters, and the transfer rate of the single conformation.} The conformation was randomly selected as the 2071 frame from the MD simulation and calculated in the QM/MM framework, from which the QM energies were extracted and listed here. The calculations were conducted by empirical HEA on a 13-qubit superconducting quantum hardware. The single-point energies are given in Hartree. The 6-31G basis set and the tPBE functional were used in all calculations. Explicit single-point calculations and Marcus parameters of ROUCCSD and noiseless HEA can be found in Supplementary Information.}
	\label{tab:2071}
	\end{table}
	
	This result compared favorably with noiseless calculations: $1.89 \times 10^8 s^{-1}$ (HEA) and $2.49 \times 10^8 s^{-1}$ (ROUCCSD), demonstrating that quantum hardware may conduct electron transfer evaluations within a compact active space despite inherent noise. More explicit single-point energies and derived parameters are detailed in the Supplementary Information.
	
	\subsection{Error analysis}
	
	The successful quantum hardware validation demonstrated current feasibility, but might raise questions about the reliability of noisy quantum devices for the electron transfer process. To address these concerns, we conducted comprehensive error analysis comparing single-point energies and derived Marcus parameters across different computational approaches.
	
	\subsubsection{Systematic error in single-point energies}
	
	\begin{figure*}[t]
		\centering
		\includegraphics[width=1\textwidth]{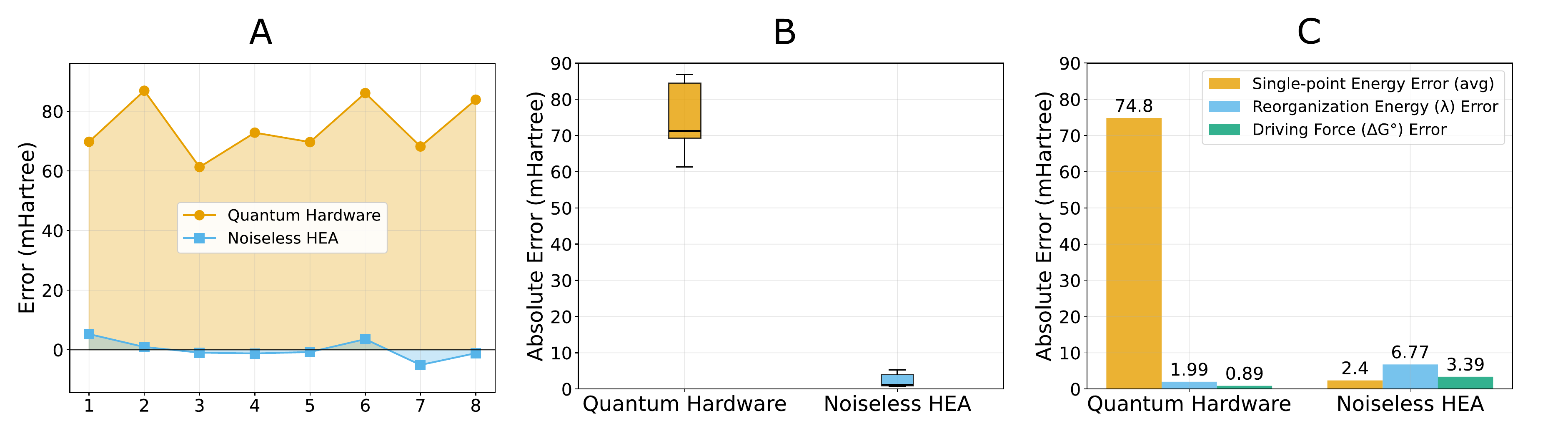}
		\caption{\textbf{Error analysis of single-point energies and Marcus parameters.} ROUCCSD's results are set as the comparing baseline. \textbf{A} Error comparison on 8 single-point energies of quantum hardware and noiseless empirical HEA, with quantum hardware's results showing systematic errors (65.3 ± 14.2 mHartree positive deviation) compared to noiseless HEA relative to ROUCCSD reference. The system indices follow the index column in Table \ref{tab:2071}. \textbf{B} Error distributions of quantum hardware and noiseless empirical HEA. \textbf{C} Error Magnitude comparison between single-point energies and Marcus parameters. Explicit single-point calculation can be found in Supplementary Information.}
		\label{fig:err_analysis}
	\end{figure*}
	
	To assess the computational accuracy of quantum hardware implementation, we examined single-point energy errors of the protein conformation (frame 2071) presented in Table \ref{tab:2071}.
	Our analysis employed ROUCCSD calculations as the reference standard, building upon our earlier validation that demonstrated excellent agreement between ROUCCSD and FCI results (Table \ref{tab:hea}).
	The quantum hardware calculations revealed systematic deviations from this ROUCCSD reference, as illustrated in Fig. \ref{fig:err_analysis} A and Fig. \ref{fig:err_analysis} B, where all eight energy calculations exhibit consistent positive shifts.
	
	This systematic bias indicated a non-random error source. The errors likely originated from correlated uncertainties in noisy matrix element measurements or from VQE-PDFT's systematic response to noisy RDMs, rather than from purely random quantum fluctuations.
	
	Moreover, the systematic nature of these hardware errors is noteworthy for biological electron transfer applications. Unlike random noise that would propagate unpredictably and through energy difference calculations, systematic errors might enable partial cancellation when computing the energy differences that underlie Marcus theory parameters. This observation motivated examining how these correlated single-point energy errors influence the derived reorganization energy and driving force that determine electron transfer rates.
	
	\subsubsection{Error cancellation in Marcus parameters}
	
	The feasibility of Marcus parameter calculations on quantum hardware can be understood through their mathematical structure. Both the driving force and reorganization energy depend on energy differences, as defined in Equations \eqref{eq:deltag} and \eqref{eq:lambda}. Consider the driving force calculation on quantum hardware:
	\begin{align}
		\Delta G^{o'} &= \left|E_{i}^{i'} - E_{f}^{f'}\right|, \nonumber\\
		&= \left|(E_{i}^{i}+\delta E_i) - (E_{f}^{f}+\delta E_f)\right|, \nonumber\\
		&= \left|(E_{i}^{i} - E_{f}^{f}) + (\delta E_i - \delta E_f)\right|,
	\end{align}
	where systematic errors $\delta E$ relative to the ROUCCSD reference partially cancel in the energy difference calculation, reducing the overall uncertainty to the net difference $(\delta E_i - \delta E_f)$. A similar partial cancellation also functions in the reorganization energy calculation. 
	
	This error cancellation effect was demonstrated in our calculations, as illustrated in Fig. \ref{fig:err_analysis} B and Fig. \ref{fig:err_analysis} C. While individual single-point energies exhibited errors of 61-86 mHartree on quantum hardware, the derived Marcus parameters showed notable reduced absolute uncertainties: $\lambda$ differed from the ROUCCSD reference by 2.0 mHartree, and $\Delta G^o$ by 0.9 mHartree. However, it should be noted that the relative errors of Marcus parameters actually increased due to their smaller absolute magnitudes ($\sim 12\%$ for $\lambda$, $\sim 36\%$ for $\Delta G^o$).
	
	Despite these relative error increases, the systematic nature of quantum hardware errors enables partial cancellation when computing energy differences. This error cancellation allowed the derived electron transfer rates ($2.62 \times 10^8 s^{-1}$) to achieve consistency with classical predictions ($1.89 \sim 2.49 \times 10^8 s^{-1}$), demonstrating that the approach remains viable for the current electron transfer application. Additionally, together with the similar effect shown in the CT7/04 benchmark test, the observed error cancellation in energy differences may thus provides a strategic pathway for NISQ devices to contribute meaningfully to computational biochemistry, by focusing on difference-based observables rather than absolute electronic energies.
	
	\section{Conclusion}\label{conclusion}
	
	In this work, we developed and validated VQE-PDFT, a quantum-classical hybrid framework that integrates variational quantum eigensolver with multiconfiguration pair-density functional theory for strongly correlated electronic systems. In VQE-PDFT, the quantum circuit is used as a CASCI active-space solver to optimize the multiconfigurational wavefunction and to sample RDMs, while the total correlation energy is evaluated classically via the MC-PDFT on-top functional as a post-processing step. This strategy delivers accuracy comparable to conventional MC-PDFT on our benchmarks, while reducing quantum resource requirements (qubit count and circuit depth) relative to highly expressive VQE approaches that seek to capture full dynamic correlation variationally by enlarging the orbital space.
	
	Validation on the CT7/04 dataset demonstrated excellent agreement with reference calculations (MUE = 0.853 kcal/mol with respect to W-1 theory), confirming the method's reliability for multiconfigurational systems. We also extended this framework to a biological electron transfer application by developing specialized hardware-efficient ansatz circuits optimized for NISQ device constraints, achieving circuit depths of 4-6, compared to 35 for ROUCCSD while maintaining high accuracy.
	
	Our comprehensive application to electron transfer in the European robin cryptochrome ErCRY4 protein yielded transfer rates ($0.944 \times 10^{10} s^{-1}$ for HEA and $0.864 \times 10^{10} s^{-1}$ for ROUCCSD) that aligned well with experimental ultrafast spectroscopy measurements ($0.709 \times 10^{10} s^{-1}$), validating our quantum-classical approach for this complex biological environment. Finally, as a proof-of-concept hardware validation within this quantum–classical multiscale framework, we executed the reduced-density-matrix measurements for a single ErCRY4 protein conformation on a 13-qubit superconducting device and obtained an estimated electron transfer rate ($2.62 \times 10^8\ \mathrm{s^{-1}}$) that aligned well with noiseless simulations, even in the presence of hardware noise.
	
	Moreover, this work also mitigates three challenges for NISQ-era quantum biology applications: explicit treatment of multiconfigurational electronic correlations through purpose-designed shallow-depth quantum circuits, enhancing the ability of calculating strongly correlated system; systematic error cancellation through focus on energy differences rather than absolute energies, where we discovered that errors partially cancel in Marcus parameter calculations with quantum hardware, alleviating the quantum noise's impact on derived observables; and integration within a scalable multiscale framework adaptable to future hardware improvements, minimizing software and algorithmic modification while maintaining utility via the flexible VQE calculation.
	
	Several limitations warrant future investigation. Firstly, current quantum hardware constraints may restrict active space sizes and computational accuracy. Secondly, efficient optimization and measurement grouping methods should be developed to lower the costs of quantum computation, enabling sampling across more protein conformations on superconducting quantum hardware, which was the main obstacle while conducting experimental validation. Thirdly, our separate treatment of indole rings on two tryptophan residues, while justified by weak electronic coupling, represents an approximation that could be improved by simultaneous evaluation of multiple rings to capture longer-range interactions and intermediate transfer states. Besides, in this proof-of-principle study we did not perform a systematic sensitivity analysis with respect to the precise QM/MM partitioning or to enlarging the tryptophan active spaces, as such an investigation would require a substantially larger set of high-level quantum calculations; exploring these alternatives will be an important direction for future work.
	
	Additionally, the empirical HEA circuits adopted here are constructed for the present compact tryptophan active spaces under fixed $(N_\alpha,N_\beta)$ constraints and are not expected to be directly transferable. Therefore, developing more systematic and automated strategies for symmetry- and configuration-guided HEA constructions, and assessing their performance in larger active spaces and broader scenarios (e.g., bond-dissociation/potential-energy curves), would also be a valuable direction for future work.
	
	Future developments will further focus on constraining electronic populations during VQE optimization through penalty functions or Lagrangian multipliers with information from Mulliken population analysis, enabling treatment of multiple simultaneous transfer pathways. As quantum hardware develops, our scalable framework can accommodate increased qubit counts and improved fidelities for more comprehensive biological system calculations.
	
	In conclusion, VQE-PDFT illustrates a practical pathway for quantum computing applications in biochemistry, suggesting that despite current limitations, quantum-classical hybrid approaches may yield reasonably accurate predictions for simplified active-space models of particular biological processes and can serve as an initial step toward exploring quantum utility in molecular sciences.
	
	\section{Methods}
	
	\subsection{Quantum computing reduced density matrices}
	
	In VQE-PDFT, VQE is used as a CASCI solver to optimize the CASCI active-space Hamiltonian, which is performed before the PDFT energy evaluation. Each VQE run is initialized in a product state consistent with the chosen ansatz (Hartree–Fock determinant for UCCSD/ROUCCSD and the all-zero computational state for the HEA circuits.) 
	Once the optimization has converged, the parameters of circuits are frozen and passed for subsequently evaluating 1-RDM and 2-RDM. 
	The matrix elements of 1-RDM and 2-RDM may be written as the expectation value in terms of creation/annihilation operators under the second quantization, such as for 1-RDM,
	$$\gamma_{pq} = \left\langle \psi \right| a^\dagger_p a_q \left| \psi \right\rangle,$$
	where the electronic state $|\psi\rangle $ can be efficiently reconstructed with the previously frozen circuit parameters. Therefore, 
	$$\gamma_{pq} = \left\langle \psi(\theta) \right| a^\dagger_p a_q \left| \psi(\theta) \right\rangle,$$
	where
	$$\left| \psi(\theta) \right\rangle = U(\theta)\left|\psi_0\right\rangle.$$
	This enables us to compute each elements of 1-RDM by executing the circuit with frozen parameters and sampling the expectation value of the corresponding operators after fermion-to-qubit mapping such as the parity transformation.
	
	The same procedure is also valid for computing 2-RDM $\Gamma_{pq,rs}$.
	
	In this element-wise measurement protocol, the additional post-optimization measurement overhead is governed by the number of distinct RDM elements to be sampled. For an active space comprising $N_\mathrm{so}$ spin orbitals, the 1-RDM $\gamma_{pq}$ and 2-RDM $\Gamma_{pq,rs}$ would formally contain $O(N_\mathrm{so}^2)$ and $O(N_\mathrm{so}^4)$ matrix elements, respectively (up to constant prefactors from Hermiticity and index symmetries). After fermion-to-qubit transformation, each fermionic operator is mapped to a constant number of Pauli words to be measured, so the number of distinct measurement operators required for sampling the full 1- and 2-RDM scales as $O(N_\mathrm{so}^2)$ for the 1-RDM and $O(N_\mathrm{so}^4)$ for the 2-RDM, respectively.
	
	Given this measurement overhead, we note that alternative strategies have been proposed to reduce the quantum measurement cost of RDM evaluations. For instance, natural orbital functional (NOF) methods reconstruct an approximate 2-RDM from measured 1-RDM, reducing formal sampling requirements from $O(N_\mathrm{so}^4)$ to $O(N_\mathrm{so}^2)$ \cite{lew-yee2025}. However, because MC-PDFT depends explicitly on 2-RDM information through the on-top pair density $\Pi$ introducing an additional 2-RDM reconstruction approximation may bias the functional evaluation. Our approach therefore directly measures the 1-RDM and 2-RDM elements in the present work. Separately, for evaluating the CASCI energy one may further reduce measurement cost by truncating Pauli terms with small Hamiltonian coefficients. This term-selection techniques are complementary to VQE-PDFT and could be combined with our framework.
	
	\subsection{Designing the empirical HEA}\label{EHEA method}
	
	The hardware-efficient ansatz (HEA) circuits are designed based on the electronic structure characteristics of tryptophan indole systems, optimizing quantum resources while maintaining computational accuracy for NISQ devices.
	For clarity, the empirical HEA construction follows a simple workflow:
	\begin{itemize}
			\item identify the active space and fermion-to-qubit mapping (Parity transformation here), and determine the fixed $N_{\alpha}$/$N_{\beta}$ symmetry sector;
			\item reduce the qubit register by tapering qubits associated with these symmetries;
			\item determine the set of computational-basis configurations that satisfy particle conservation;
			\item construct a shallow symmetry-preserving circuit that couples these configurations with minimal depth and parameter count.
	\end{itemize}
	
	\subsubsection*{Qubit reduction through symmetry constraints}
	The (4e,3o) closed-shell and (3e,3o) open-shell active spaces require 6 qubits under parity transformation. Exploiting $\alpha$ and $\beta$ electron number conservation reduces this to 4 qubits, as qubits 3 and 6 (representing total $\alpha$-electron and total electron occupation parity) remain fixed throughout calculations. This symmetry-based tapering reduces quantum resource requirements while preserving all physically accessible states.
	
	\subsubsection*{Determine the symmetry-allowed configuration subspace}
	After tapering, particle-number conservation further restricts the computational basis allowed to access in the reduced qubit register. For example, in open-shell (3e,3o) systems with 2$\alpha$ and 1$\beta$ electrons, the first two qubits must not access $|00\rangle$ (violating $\alpha$-electron conservation) while qubits 3 and 4 must not access $|10\rangle$ (violating $\beta$-electron conservation), restricting the total accessible space to 3×3=9 states. Consistent with this symmetry restriction, the converged ROUCCSD wavefunctions have support only within this nine-state sector for the present active spaces.
	
	\subsubsection*{Circuit construction based on accessible subspace}
	With the symmetry constraints enforced and the accessible configuration subspace identified, we construct the CHEA/OHEA circuits in a pragmatic, empirical manner tailored to the present compact active spaces for the proof-of-concept demonstration. In addition, we may even select the highest-weight configurations within the symmetry-allowed subspace as the dominant configurations to be captured, via inspecting classical reference in the same active space (such as ROUCCSD), and further reduce the target subspace. Then, we iteratively assemble a shallow circuit from a restricted gate set \{$\mathrm{X}, \mathrm{R}_y, \mathrm{CNOT}, \mathrm{CR}_y$\}: at each step, candidate gate additions are evaluated classically on the reduced register to ensure the resulting state maintains particle number conservation and to increase the circuit’s ability to populate the target subspace.
		
	Owing to the strong particle-number constraints and the small, factorized structure of the $\alpha$- and $\beta$-spin qubit registers in the present (4e,3o)/(3e,3o) cases, this iterative construction remains tractable and yields depth-4 (CHEA, Fig. \ref{fig:hea} A) and depth-6 (OHEA, Fig. \ref{fig:hea} B) circuits that are sufficiently flexible for the subsequent validation and hardware execution, compared to ROUCCSD's 35-layer requirement. 
	A quantitative comparison of the logical circuit depths, total numbers of single- and two-qubit gates for ROUCCSD and the HEA circuits is provided in Supplementary Table S2(a).
	
	This empirical approach represents a compromise necessitated by current NISQ limitations rather than fundamental algorithmic constraints that can be directly transferred for arbitrary systems. 
	The system-specific ansatz optimized for tryptophan electronic structure patterns requires extensive reanalysis for different chemical environments or larger active spaces. 
	For other systems, the same design workflow may be followed, but the resulting circuit structure will generally change with (a) the active-space size and occupation pattern (hence the number of qubits after symmetry tapering) and (b) the number and connectivity of dominant configurations within the target symmetry sector, which may require additional entangling layers and variational parameters to maintain accuracy.
	As quantum hardware develops to support deeper circuits (hundreds of circuit-depths) with high fidelity, more general approaches like ROUCCSD or highly expressive multi-layer HEAs may become preferable over these tailored solutions \cite{guo2024}. 
	
	\subsection{Computational details}
	
	\subsubsection*{Quantum hardware implementation}
	VQE optimizations were performed on classical simulators to avoid noise accumulation during iterative parameter updates. Once parameter convergence was achieved, the optimized parameters were saved and transferred to quantum hardware for reduced density matrix evaluations. 
	
\begin{figure}[htb]
		\centering
		\includegraphics[width=0.3\textwidth]{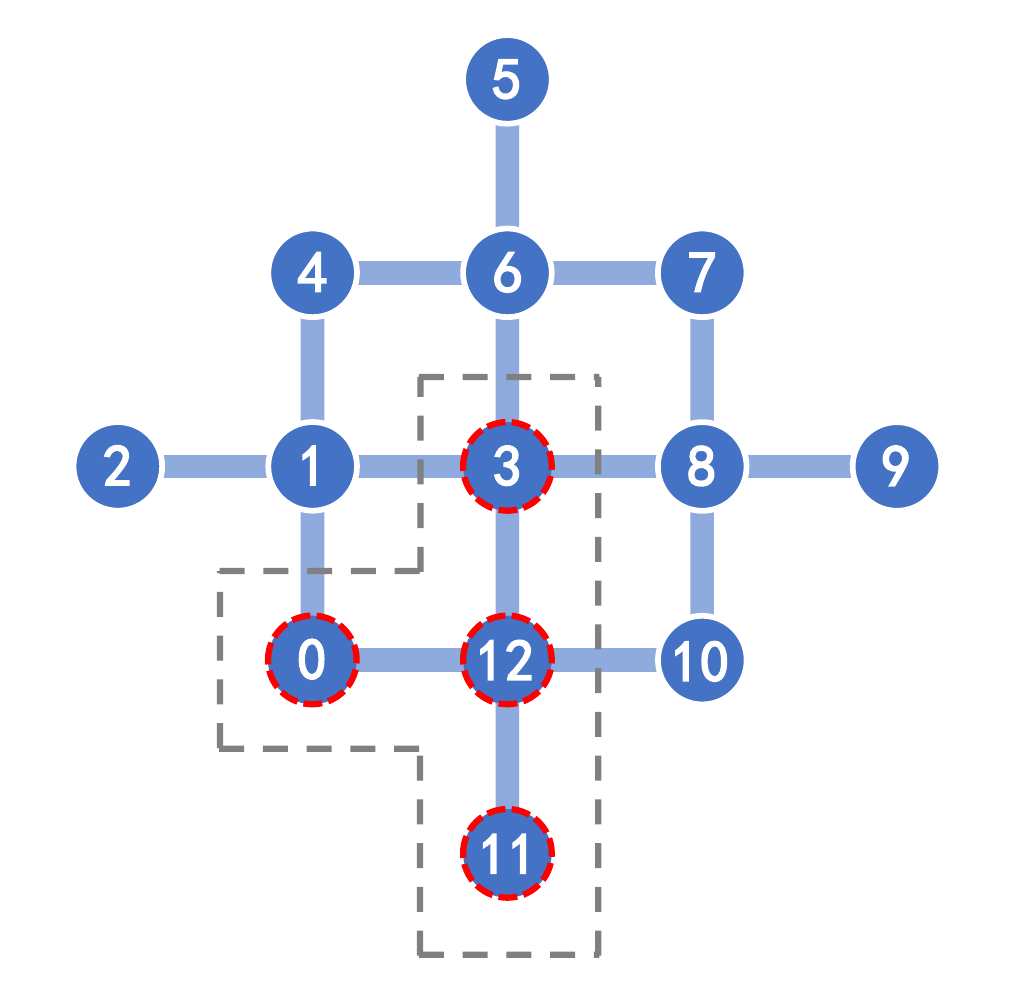}
		\caption{\textbf{Topology of 13-qubit superconducting quantum processor.} During our validations, we only used 4 qubits indexing \{0, 3, 11, 12\} and their couplings.}
		\label{fig:top}
	\end{figure}

	\noindent\textbf{Hardware specification.} Quantum computation employed a customized 13-qubit superconducting quantum device with the following performance metrics: single-qubit gate fidelity of 99.93\%, two-qubit CZ gate fidelity of 99.13\%, coherence times of T\textsubscript{1} = 83.8 $\mu s$ and T\textsubscript{2} = 45.6 $\mu s$, and readout fidelities of F\textsubscript{0} = 98.35\% and F\textsubscript{1} = 95.88\%. The connecting topology is shown in Fig. \ref{fig:top}.
	
	\noindent\textbf{RDM matrix element evaluation.} The computation of 1-RDM and 2-RDM matrix elements followed a systematic six-step protocol:
	\begin{enumerate}
		\item \textbf{\textit{Pauli operators preparation}}. Measurement Pauli operators were obtained through parity transformation of the corresponding matrix elements.
		\item \textbf{\textit{Measurement grouping}}. Compatible measurement operators were grouped using a simple grouping strategy to reduce quantum measurement overhead and noise impact. For instance, operators $Z_1Z_2I_3I_4$ and $I_1I_2Z_3Z_4$ can be simultaneously measured using the single operation $Z_1Z_2Z_3Z_4$.
		\item \textbf{\textit{Circuit compilation}}. The empirical HEA circuits and grouped measurements were uploaded to the quantum hardware platform. These circuits then underwent qubit mapping and gate decomposition to match the chip topology and native gate set. As shown in Fig. \ref{fig:top}, the resulting compiled circuits utilized only 4 physical qubits, benefiting from our restrained use of two-qubit gates that eliminated the need for additional connectivity qubits. Quantitative information for the compiled HEA circuits on the 13-qubit device are summarized in Supplementary Table S2(b).
		\item \textbf{\textit{Circuit execution}}. Each compiled circuit was executed with 2048 measurement shots, returning raw bit-string distribution data.
		\item \textbf{\textit{Readout error mitigation}}. Raw measurement data underwent readout error correction using the correlated Markovian noise model approach. This correction required calibration circuits executed once per single-point energy calculation with 8192 shots. The resulting calibration data was then stored for all subsequent measurements within that calculation. Finally, the error mitigation module processed raw bit-string distribution, calibration data, and target measurement operators to yield the corrected expectation value as the matrix elements.
		\item \textbf{\textit{RDMs assembly}}. Steps 1-5 were repeated for all required matrix elements to construct the entire 1-RDM and 2-RDM matrices.
	\end{enumerate}
	This protocol enabled reliable extraction of reduced density matrices from noisy quantum hardware while maintaining computational efficiency through strategic measurement grouping and error mitigation.
	
	\subsubsection*{Molecular dynamics sampling and statistical analysis}
	Twenty protein conformations were randomly sampled from molecular dynamics simulations of ErCRY4 \cite{xu2021}. Every indices of these conformations are detailed in Supplementary Information. Marcus theory parameters (reorganization energy $\lambda$, driving force $\Delta G^o$, and electronic coupling $\left\langle |H_{DA}|\right\rangle $) were calculated independently for each conformation, with statistical averages of these parameters input into the Marcus rate expression. This approach ensures thermodynamically consistent averaging of Marcus theory parameters, representing the actual biological environment where conformational fluctuations modulate the underlying electronic structure properties.
	
	\subsubsection*{Software implementation}
	The VQE-PDFT framework was implemented in Python through local modifications of PySCF \cite{sun2020} and TenCirChem \cite{li2023c} libraries. QM/MM calculations employed the ASH \cite{bjornsson} framework for the link-atom and electrostatic embedding between the QM regions and the MM regions. Quantum hardware validation process was performed via the TensorCircuit \cite{zhang2023b} library for measurement grouping, communicating quantum device, and the readout error mitigation.
	
	\section*{Data and code availability}
	The CT7/04 dimers reported in Ref. \cite{zhao2005,zhao2005a} can be fetched from \url{https://comp.chem.umn.edu/db/dbs/ncce31.html}. The MD simulation file of ErCRY4 reported in Ref. \cite{xu2021} can be obtained from \url{https://cloud.uol.de/s/NrTYpoEzL6RbPq7}. Other relevant data supporting this study are available from the corresponding author upon reasonable request. The custom code and local modifications for implementation are available in the GitHub repository: \url{https://github.com/yiboch/VQE-PDFT}.
	
	\section*{Acknowledgments}
	This work is supported by State Key Laboratory of Genome and Multi-omics Technologies, the Guangdong Bigdata Engineering Technology Research Center for Life Sciences and the Shenzhen City Peacock Team Project (grant no. KQTD20240729102028011 to W.L.).
	
	\section*{Declaration of interests}
	The authors declare no competing interests.
	
	\section*{Author Contributions}
	Conceptualization: Y.C., and W.L.; Methodology: Y.C., Z.S., and W.L.; Software: Y.C., Z.S., and W.L.; Validation: Y.C., Z.S., and W.L.; Formal analysis: Y.C., and W.L.; Investigation: Y.C.; Resources: J.-H.H., and Y.L.; Data curation: Y.C.; Writing - original draft: Y.C., and Z.S.; Writing - review \& editing: Y.C., Z.S., W.L., Y.Z., X.X., J.-H.H., and Y.L.; Visualization: Y.C.; Supervision: W.L., Y.Z., X.X., J.-H.H., and Y.L.; Project administration: J.-H.H.; Funding acquisition: W.L., J.-H.H., Y.Z., and Y.L.

\bibliographystyle{vancouver}
\bibliography{ref}

\begin{thebibliography}{10}

\bibitem{yanagisawa2019}
Yanagisawa T.
\newblock Mechanism of High-Temperature Superconductivity in
  Correlated-Electron Systems.
\newblock Condensed Matter. 2019;4(2):57.

\bibitem{biz2021}
Biz C, Fianchini M, Gracia J.
\newblock Strongly Correlated Electrons in Catalysis: Focus on Quantum
  Exchange.
\newblock ACS Catalysis. 2021;11(22):14249-61.

\bibitem{reiher2017}
Reiher M, Wiebe N, Svore KM, Wecker D, Troyer M.
\newblock Elucidating Reaction Mechanisms on Quantum Computers.
\newblock Proceedings of the National Academy of Sciences.
  2017;114(29):7555-60.

\bibitem{zhang2015}
Zhang Y, Berman GP, Kais S.
\newblock The Radical Pair Mechanism and the Avian Chemical Compass: Quantum
  Coherence and Entanglement.
\newblock International Journal of Quantum Chemistry. 2015;115(19):1327-41.

\bibitem{xu2021}
Xu J, Jarocha LE, Zollitsch T, Konowalczyk M, Henbest KB, Richert S, et~al.
\newblock Magnetic Sensitivity of Cryptochrome 4 from a Migratory Songbird.
\newblock Nature. 2021;594(7864):535-40.

\bibitem{zhou2022}
Zhou C, Hermes MR, Wu D, Bao JJ, Pandharkar R, King DS, et~al.
\newblock Electronic Structure of Strongly Correlated Systems: Recent
  Developments in Multiconfiguration Pair-Density Functional Theory and
  Multiconfiguration Nonclassical-Energy Functional Theory.
\newblock Chemical Science. 2022;13(26):7685-706.

\bibitem{levine2021}
Levine BG, Durden AS, Esch MP, Liang F, Shu Y.
\newblock CAS without SCF—Why to Use CASCI and Where to Get the Orbitals.
\newblock The Journal of Chemical Physics. 2021;154(9).

\bibitem{olsen2011}
Olsen J.
\newblock The CASSCF Method: A Perspective and Commentary.
\newblock International Journal of Quantum Chemistry. 2011;111(13):3267-72.

\bibitem{wallace2014}
Wallace AJ, Crittenden DL.
\newblock Optimal Composition of Atomic Orbital Basis Sets for Recovering
  Static Correlation Energies.
\newblock The Journal of Physical Chemistry A. 2014;118(11):2138-48.

\bibitem{pathak2017}
Pathak S, Lang L, Neese F.
\newblock A Dynamic Correlation Dressed Complete Active Space Method: Theory,
  Implementation, and Preliminary Applications.
\newblock The Journal of Chemical Physics. 2017;147(23).

\bibitem{benavides-riveros2017}
Benavides-Riveros CL, Lathiotakis NN, Marques MAL.
\newblock Towards a Formal Definition of Static and Dynamic Electronic
  Correlations.
\newblock Physical Chemistry Chemical Physics. 2017;19(20):12655-64.

\bibitem{helgaker2000}
Helgaker T, Jørgensen P, Olsen J.
\newblock In: Multiconfigurational Self‐Consistent Field Theory. 1st ed.;
  2000. p. 598-647.

\bibitem{roos1996}
Roos BO, Andersson K, Fülscher MP, Malmqvist P, Serrano‐Andrés L, Pierloot
  K, et~al.
\newblock In: Multiconfigurational Perturbation Theory: Applications in
  Electronic Spectroscopy. 1st ed.; 1996. p. 219-331.

\bibitem{battaglia2023}
Battaglia S, Fdez~Galván I, Lindh R.
\newblock In: Multiconfigurational Quantum Chemistry: The CASPT2 Method; 2023.
  p. 135-62.

\bibitem{limanni2014}
Li~Manni G, Carlson RK, Luo S, Ma D, Olsen J, Truhlar DG, et~al.
\newblock Multiconfiguration Pair-Density Functional Theory.
\newblock Journal of Chemical Theory and Computation. 2014;10(9):3669-80.

\bibitem{ghosh2015}
Ghosh S, Sonnenberger AL, Hoyer CE, Truhlar DG, Gagliardi L.
\newblock Multiconfiguration Pair-Density Functional Theory Outperforms
  Kohn–Sham Density Functional Theory and Multireference Perturbation Theory
  for Ground-State and Excited-State Charge Transfer.
\newblock Journal of Chemical Theory and Computation. 2015;11(8):3643-9.

\bibitem{gagliardi2017}
Gagliardi L, Truhlar DG, Li~Manni G, Carlson RK, Hoyer CE, Bao JL.
\newblock Multiconfiguration Pair-Density Functional Theory: A New Way To Treat
  Strongly Correlated Systems.
\newblock Accounts of Chemical Research. 2017;50(1):66-73.

\bibitem{levine2020}
Levine DS, Hait D, Tubman NM, Lehtola S, Whaley KB, Head-Gordon M.
\newblock CASSCF with Extremely Large Active Spaces Using the Adaptive Sampling
  Configuration Interaction Method.
\newblock Journal of Chemical Theory and Computation. 2020;16(4):2340-54.

\bibitem{kandala2017}
Kandala A, Mezzacapo A, Temme K, Takita M, Brink M, Chow JM, et~al.
\newblock Hardware-Efficient Variational Quantum Eigensolver for Small
  Molecules and Quantum Magnets.
\newblock Nature. 2017;549(7671):242-6.

\bibitem{googleaiquantumandcollaborators2020}
{GOOGLE AI QUANTUM AND COLLABORATORS}, Arute F, Arya K, Babbush R, Bacon D,
  Bardin JC, et~al.
\newblock Hartree-Fock on a Superconducting Qubit Quantum Computer.
\newblock Science. 2020;369(6507):1084-9.

\bibitem{tazhigulov2022}
Tazhigulov RN, Sun SN, Haghshenas R, Zhai H, Tan ATK, Rubin NC, et~al.
\newblock Simulating Models of Challenging Correlated Molecules and Materials
  on the Sycamore Quantum Processor.
\newblock PRX Quantum. 2022;3(4).

\bibitem{preskill2018}
Preskill J.
\newblock Quantum Computing in the NISQ Era and Beyond.
\newblock Quantum. 2018;2:79.

\bibitem{peruzzo2014}
Peruzzo A, McClean J, Shadbolt P, Yung MH, Zhou XQ, Love PJ, et~al.
\newblock A Variational Eigenvalue Solver on a Photonic Quantum Processor.
\newblock Nature Communications. 2014;5(1).

\bibitem{grimsley2019}
Grimsley HR, Economou SE, Barnes E, Mayhall NJ.
\newblock An Adaptive Variational Algorithm for Exact Molecular Simulations on
  a Quantum Computer.
\newblock Nature Communications. 2019;10(1):3007.

\bibitem{tang2021}
Tang HL, Shkolnikov VO, Barron GS, Grimsley HR, Mayhall NJ, Barnes E, et~al.
\newblock Qubit-ADAPT-VQE: An Adaptive Algorithm for Constructing
  Hardware-Efficient Ansatze on a Quantum Processor.
\newblock PRX Quantum. 2021;2(2):020310.

\bibitem{zhang2022d}
Zhang Y, Cincio L, Negre CFA, Czarnik P, Coles PJ, Anisimov PM, et~al.
\newblock Variational Quantum Eigensolver with Reduced Circuit Complexity.
\newblock npj Quantum Information. 2022;8(1):1-10.

\bibitem{li2022}
Li W, Huang Z, Cao C, Huang Y, Shuai Z, Sun X, et~al.
\newblock Toward Practical Quantum Embedding Simulation of Realistic Chemical
  Systems on Near-Term Quantum Computers.
\newblock Chemical Science. 2022;13(31):8953-62.

\bibitem{anselmetti2021}
Anselmetti GLR, Wierichs D, Gogolin C, Parrish RM.
\newblock Local, Expressive, Quantum-Number-Preserving VQE Ansätze for
  Fermionic Systems.
\newblock New Journal of Physics. 2021;23(11):113010.

\bibitem{senjean2023}
Senjean B, Yalouz S, Saubanère M.
\newblock Toward Density Functional Theory on Quantum Computers?
\newblock SciPost Physics. 2023;14(3):055.

\bibitem{hardikar2024}
Hardikar TS, Heitritter K, Brown J, D'Cunha R, Mitra A, Weatherly S, et~al.
\newblock Quanta-Bind: A Quantum Computing Pipeline for Modeling Strongly
  Correlated Metal-Protein Interactions.
\newblock In: 2024 IEEE International Conference on Quantum Computing and
  Engineering (QCE); 2024. p. 538-44.

\bibitem{otten2022}
Otten M, Hermes MR, Pandharkar R, Alexeev Y, Gray SK, Gagliardi L.
\newblock Localized Quantum Chemistry on Quantum Computers.
\newblock Journal of Chemical Theory and Computation. 2022;18(12):7205-17.

\bibitem{mitra2024}
Mitra A, D’Cunha R, Wang Q, Hermes MR, Alexeev Y, Gray SK, et~al.
\newblock The Localized Active Space Method with Unitary Selective Coupled
  Cluster.
\newblock Journal of Chemical Theory and Computation. 2024:acs.jctc.4c00528.

\bibitem{sugisaki2022a}
Sugisaki K, Kato T, Minato Y, Okuwaki K, Mochizuki Y.
\newblock Variational Quantum Eigensolver Simulations with the Multireference
  Unitary Coupled Cluster Ansatz: A Case Study of the C\textsubscript{2v}
  Quasi-Reaction Pathway of Beryllium Insertion into a H\textsubscript{2}
  Molecule.
\newblock Physical Chemistry Chemical Physics. 2022;24(14):8439-52.

\bibitem{wecker2015}
Wecker D, Hastings MB, Troyer M.
\newblock Progress towards Practical Quantum Variational Algorithms.
\newblock Physical Review A. 2015;92(4):042303.

\bibitem{lee2019}
Lee J, Huggins WJ, Head-Gordon M, Whaley KB.
\newblock Generalized Unitary Coupled Cluster Wave Functions for Quantum
  Computation.
\newblock Journal of Chemical Theory and Computation. 2019;15(1):311-24.

\bibitem{dasgupta2024}
Dasgupta S, Humble T.
\newblock Impact of Unreliable Devices on Stability of Quantum Computations.
\newblock ACM Transactions on Quantum Computing. 2024;5(4):1-23.

\bibitem{lolur2023}
Lolur P, Skogh M, Dobrautz W, Warren C, Biznárová J, Osman A, et~al.
\newblock Reference-State Error Mitigation: A Strategy for High Accuracy
  Quantum Computation of Chemistry.
\newblock Journal of Chemical Theory and Computation. 2023;19(3):783-9.

\bibitem{tilly2021}
Tilly J, Sriluckshmy PV, Patel A, Fontana E, Rungger I, Grant E, et~al.
\newblock Reduced Density Matrix Sampling: Self-Consistent Embedding and
  Multiscale Electronic Structure on Current Generation Quantum Computers.
\newblock Physical Review Research. 2021;3(3):033230.

\bibitem{matousek2024}
Matoušek M, Pernal K, Pavošević F, Veis L.
\newblock Variational Quantum Eigensolver Boosted by Adiabatic Connection.
\newblock The Journal of Physical Chemistry A. 2024;128(3):687-98.

\bibitem{boyn2021}
Boyn JN, Lykhin AO, Smart SE, Gagliardi L, Mazziotti DA.
\newblock Quantum-Classical Hybrid Algorithm for the Simulation of All-Electron
  Correlation.
\newblock The Journal of Chemical Physics. 2021;155(24):244106.

\bibitem{timmer2023}
Timmer D, Frederiksen A, Lünemann DC, Thomas AR, Xu J, Bartölke R, et~al.
\newblock Tracking the Electron Transfer Cascade in European Robin Cryptochrome
  4 Mutants.
\newblock Journal of the American Chemical Society. 2023;145(21):11566-78.

\bibitem{zhao2005}
Zhao Y, Truhlar DG.
\newblock Benchmark Databases for Nonbonded Interactions and Their Use To Test
  Density Functional Theory.
\newblock Journal of Chemical Theory and Computation. 2005;1(3):415-32.

\bibitem{vitillo2022}
Vitillo JG, Cramer CJ, Gagliardi L.
\newblock Multireference Methods Are Realistic and Useful Tools for Modeling
  Catalysis.
\newblock Israel Journal of Chemistry. 2022;62(1--2).

\bibitem{lischka2018}
Lischka H, Nachtigallová D, Aquino AJA, Szalay PG, Plasser F, Machado FBC,
  et~al.
\newblock Multireference Approaches for Excited States of Molecules.
\newblock Chemical Reviews. 2018;118(15):7293-361.

\bibitem{zhao2005a}
Zhao Y, Truhlar DG.
\newblock Design of Density Functionals That Are Broadly Accurate for
  Thermochemistry, Thermochemical Kinetics, and Nonbonded Interactions.
\newblock The Journal of Physical Chemistry A. 2005;109(25):5656-67.

\bibitem{papajak2011}
Papajak E, Zheng J, Xu X, Leverentz HR, Truhlar DG.
\newblock Perspectives on Basis Sets Beautiful: Seasonal Plantings of Diffuse
  Basis Functions.
\newblock Journal of Chemical Theory and Computation. 2011;7(10):3027-34.

\bibitem{meng2023}
Meng EC, Goddard TD, Pettersen EF, Couch GS, Pearson ZJ, Morris JH, et~al.
\newblock ChimeraX.
\newblock Protein Science. 2023;32(11):e4792.

\bibitem{li2019}
Li Z, Li J, Dattani NS, Umrigar CJ, Chan GKL.
\newblock The Electronic Complexity of the Ground-State of the FeMo Cofactor of
  Nitrogenase as Relevant to Quantum Simulations.
\newblock The Journal of Chemical Physics. 2019;150(2):024302.

\bibitem{li2024a}
Li W, Yin Z, Li X, Ma D, Yi S, Zhang Z, et~al.
\newblock A Hybrid Quantum Computing Pipeline for Real World Drug Discovery.
\newblock Scientific Reports. 2024;14(1):16942.

\bibitem{senn2009}
Senn HM, Thiel W.
\newblock QM/MM Methods for Biomolecular Systems.
\newblock Angewandte Chemie International Edition. 2009;48(7):1198-229.

\bibitem{sun2020}
Sun Q, Zhang X, Banerjee S, Bao P, Barbry M, Blunt NS, et~al.
\newblock Recent Developments in the PySCF Program Package.
\newblock The Journal of Chemical Physics. 2020;153(2):024109.

\bibitem{li2023c}
Li W, Allcock J, Cheng L, Zhang SX, Chen YQ, Mailoa JP, et~al.
\newblock TenCirChem: An Efficient Quantum Computational Chemistry Package for
  the NISQ Era.
\newblock Journal of Chemical Theory and Computation. 2023;19(13):3966-81.

\bibitem{bergholm2022}
Bergholm V, Izaac J, Schuld M, Gogolin C, Ahmed S, Ajith V, et~al.
\newblock PennyLane: Automatic Differentiation of Hybrid Quantum-Classical
  Computations.
\newblock arXiv preprint arXiv:181104968. 2022.
\newblock Quant-ph.

\bibitem{lopez-estrada2018}
López-Estrada O, Laguna HG, Barrueta-Flores C, Amador-Bedolla C.
\newblock Reassessment of the Four-Point Approach to the Electron-Transfer
  Marcus–Hush Theory.
\newblock ACS Omega. 2018;3(2):2130-40.

\bibitem{marcus1985}
Marcus RA, Sutin N.
\newblock Electron Transfers in Chemistry and Biology.
\newblock Biochimica et Biophysica Acta (BBA) - Reviews on Bioenergetics.
  1985;811(3):265-322.

\bibitem{firmino2016}
Firmino T, Mangaud E, Cailliez F, Devolder A, Mendive-Tapia D, Gatti F, et~al.
\newblock Quantum Effects in Ultrafast Electron Transfers within Cryptochromes.
\newblock Physical Chemistry Chemical Physics. 2016;18(31):21442-57.

\bibitem{subotnik2008}
Subotnik JE, Yeganeh S, Cave RJ, Ratner MA.
\newblock Constructing Diabatic States from Adiabatic States: Extending
  Generalized Mulliken–Hush to Multiple Charge Centers with Boys
  Localization.
\newblock The Journal of Chemical Physics. 2008;129(24).

\bibitem{hsu2009}
Hsu CP.
\newblock The Electronic Couplings in Electron Transfer and Excitation Energy
  Transfer.
\newblock Accounts of Chemical Research. 2009;42(4):509-18.

\bibitem{bravyi2021}
Bravyi S, Sheldon S, Kandala A, Mckay DC, Gambetta JM.
\newblock Mitigating Measurement Errors in Multiqubit Experiments.
\newblock Physical Review A. 2021;103(4).

\bibitem{lew-yee2025}
Lew-Yee JFH, Piris M.
\newblock Efficient Energy Measurement of Chemical Systems via One-Particle
  Reduced Density Matrix: A NOF-VQE Approach for Optimized Sampling.
\newblock Journal of Chemical Theory and Computation. 2025;21(5):2402-13.

\bibitem{guo2024}
Guo S, Sun J, Qian H, Gong M, Zhang Y, Chen F, et~al.
\newblock Experimental Quantum Computational Chemistry with Optimized Unitary
  Coupled Cluster Ansatz.
\newblock Nature Physics. 2024;20(8):1240-6.

\bibitem{bjornsson}
Bj{\"o}rnsson R. ASH - a Multiscale Modelling Program;.
\newblock https://ash.readthedocs.io/.

\bibitem{zhang2023b}
Zhang SX, Allcock J, Wan ZQ, Liu S, Sun J, Yu H, et~al.
\newblock TensorCircuit: A Quantum Software Framework for the NISQ Era.
\newblock Quantum. 2023;7:912.

\end{thebibliography}


\begin{thebibliography}{1}

\bibitem{sun2020}
Sun Q, Zhang X, Banerjee S, Bao P, Barbry M, Blunt NS, et~al.
\newblock Recent Developments in the PySCF Program Package.
\newblock The Journal of Chemical Physics. 2020;153(2):024109.

\bibitem{meng2023}
Meng EC, Goddard TD, Pettersen EF, Couch GS, Pearson ZJ, Morris JH, et~al.
\newblock ChimeraX.
\newblock Protein Science. 2023;32(11):e4792.

\bibitem{li2023c}
Li W, Allcock J, Cheng L, Zhang SX, Chen YQ, Mailoa JP, et~al.
\newblock TenCirChem: An Efficient Quantum Computational Chemistry Package for
  the NISQ Era.
\newblock Journal of Chemical Theory and Computation. 2023;19(13):3966-81.

\bibitem{zhang2023b}
Zhang SX, Allcock J, Wan ZQ, Liu S, Sun J, Yu H, et~al.
\newblock TensorCircuit: A Quantum Software Framework for the NISQ Era.
\newblock Quantum. 2023;7:912.

\bibitem{ghosh2015}
Ghosh S, Sonnenberger AL, Hoyer CE, Truhlar DG, Gagliardi L.
\newblock Multiconfiguration Pair-Density Functional Theory Outperforms
  Kohn–Sham Density Functional Theory and Multireference Perturbation Theory
  for Ground-State and Excited-State Charge Transfer.
\newblock Journal of Chemical Theory and Computation. 2015;11(8):3643-9.

\bibitem{papajak2011}
Papajak E, Zheng J, Xu X, Leverentz HR, Truhlar DG.
\newblock Perspectives on Basis Sets Beautiful: Seasonal Plantings of Diffuse
  Basis Functions.
\newblock Journal of Chemical Theory and Computation. 2011;7(10):3027-34.

\bibitem{xu2021}
Xu J, Jarocha LE, Zollitsch T, Konowalczyk M, Henbest KB, Richert S, et~al.
\newblock Magnetic Sensitivity of Cryptochrome 4 from a Migratory Songbird.
\newblock Nature. 2021;594(7864):535-40.

\end{thebibliography}

\end{CJK}
\end{document}

\maketitle
	\section{Supplementary Figures}
	
	\begin{figure}[htbp]
		\centering
		% ==================== (open-shell) ====================
		\begin{subfigure}[b]{0.32\textwidth}
			\centering
			\includegraphics[width=\linewidth]{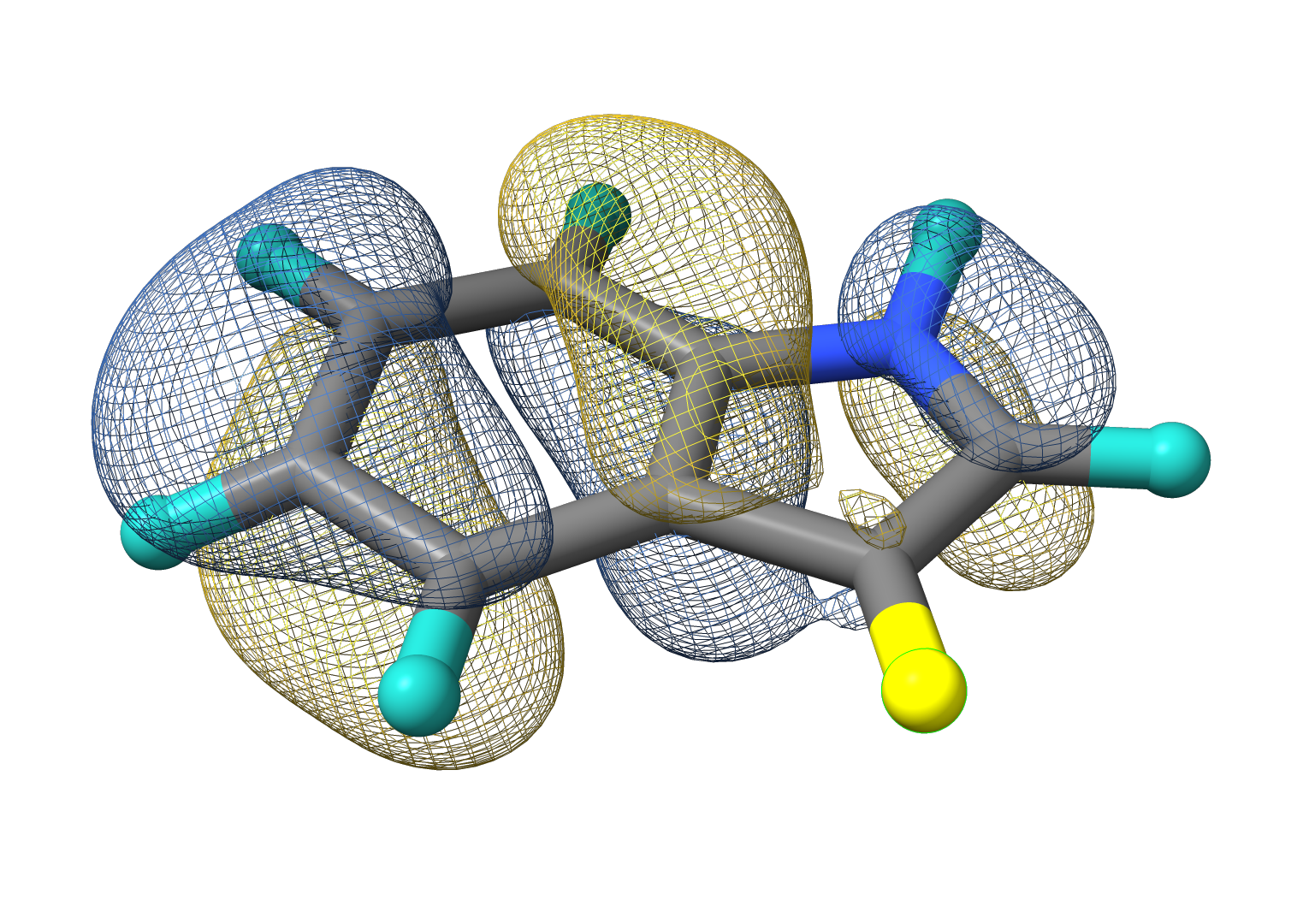} 
			\caption{\rev{open-shell active orbital 1}}
		\end{subfigure}
		\hfill
		\begin{subfigure}[b]{0.32\textwidth}
			\centering
			\includegraphics[width=\linewidth]{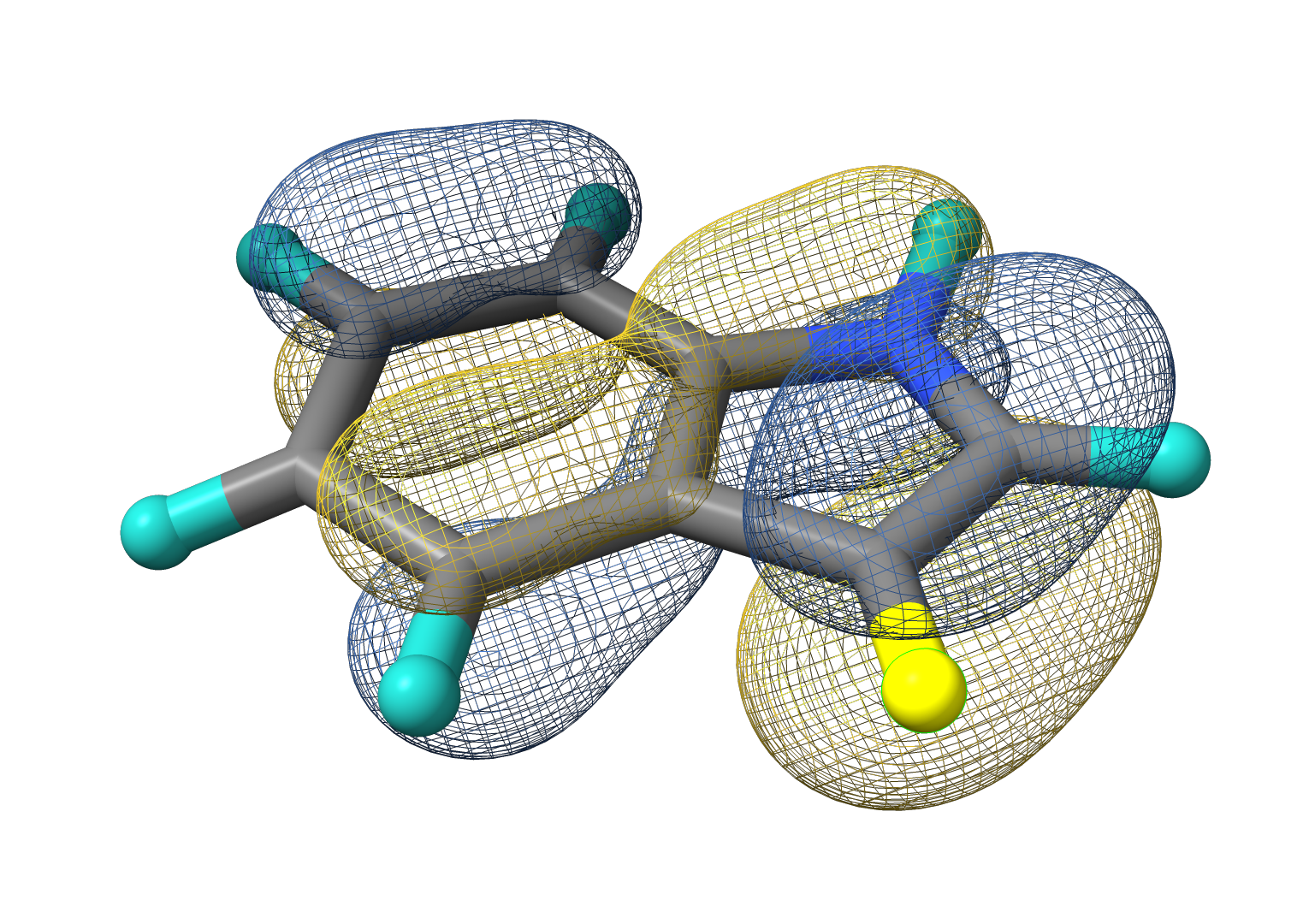}
			\caption{\rev{open-shell active orbital 2}}
		\end{subfigure}
		\hfill
		\begin{subfigure}[b]{0.32\textwidth}
			\centering
			\includegraphics[width=\linewidth]{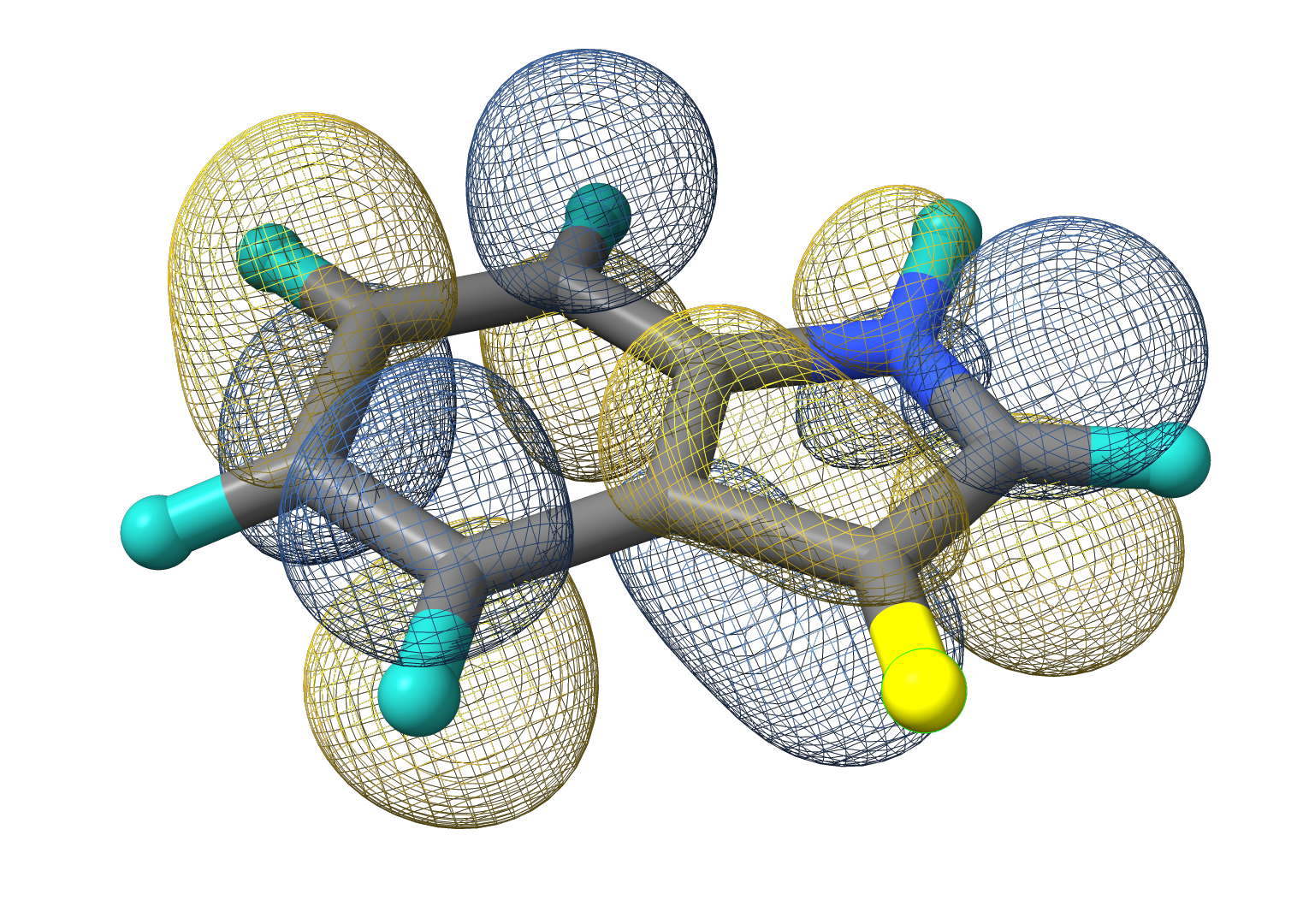}
			\caption{\rev{open-shell active orbital 3}}
		\end{subfigure}
		\vspace{0.5cm}
		% ==================== (closed shell) ====================
		\begin{subfigure}[b]{0.32\textwidth}
			\centering
			\includegraphics[width=\linewidth]{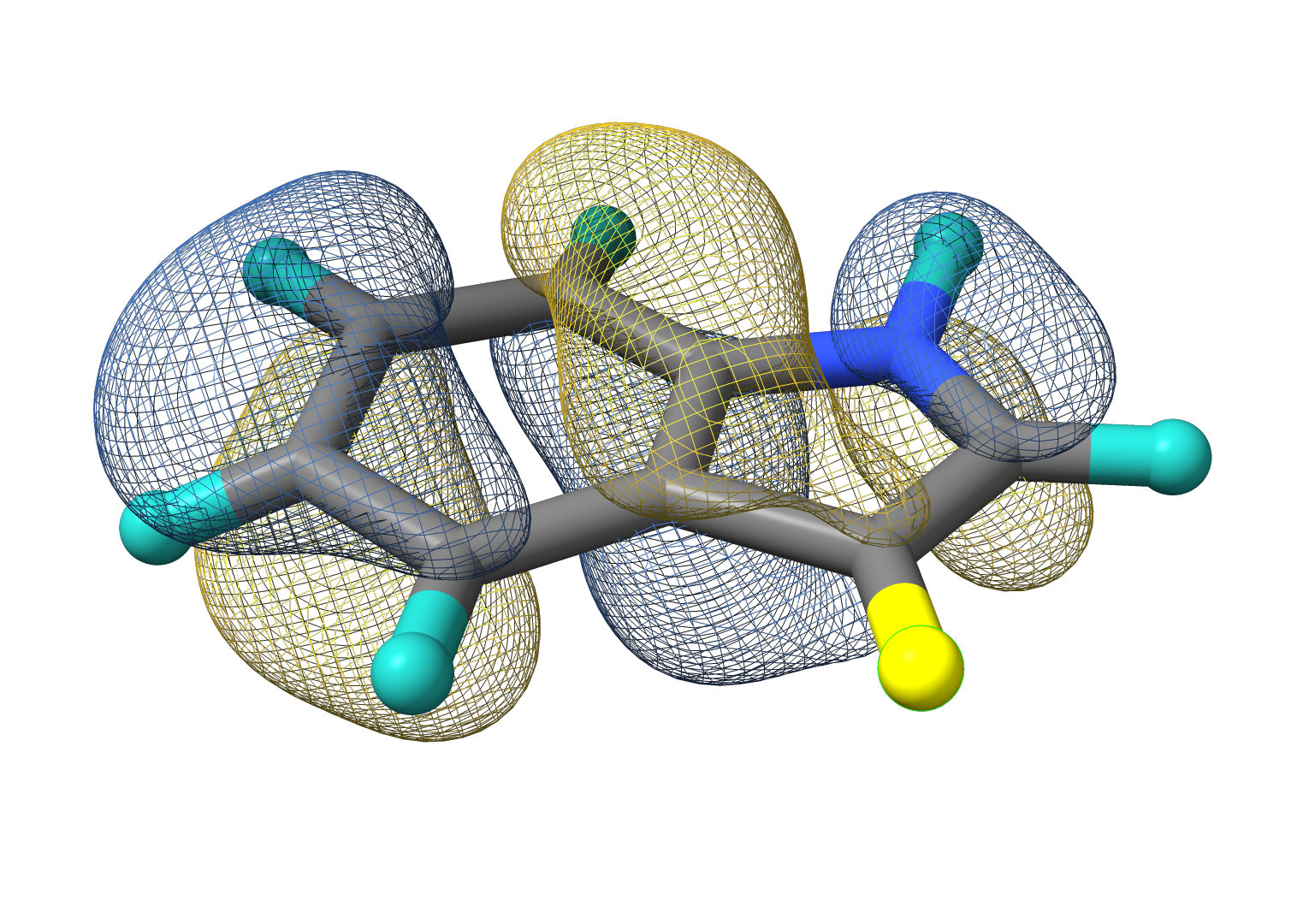}
			\caption{\rev{closed-shell active orbital 1}}
		\end{subfigure}
		\hfill
		\begin{subfigure}[b]{0.32\textwidth}
			\centering
			\includegraphics[width=\linewidth]{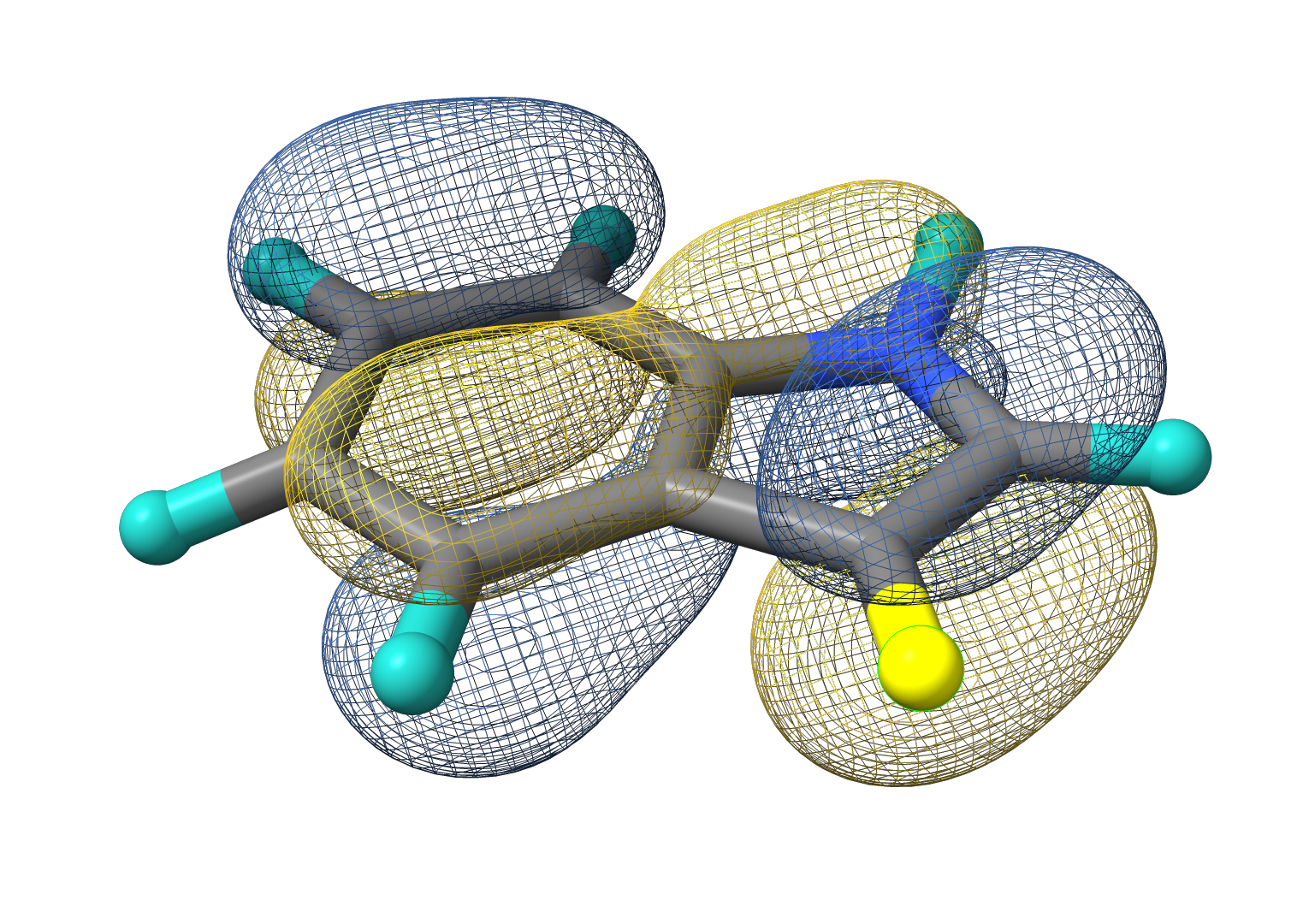}
			\caption{\rev{closed-shell active orbital 2}}
		\end{subfigure}
		\hfill
		\begin{subfigure}[b]{0.32\textwidth}
			\centering
			\includegraphics[width=\linewidth]{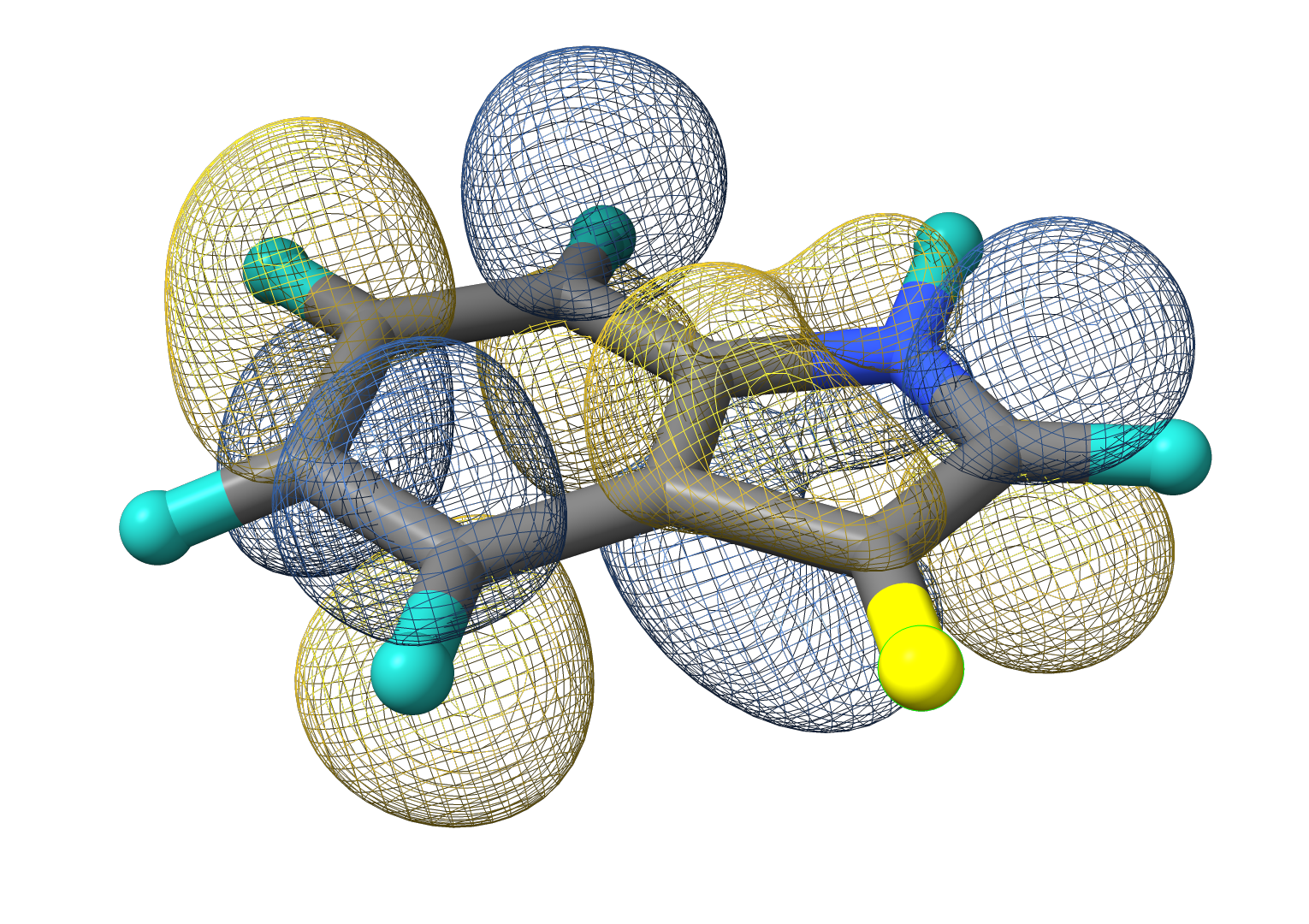}
			\caption{\rev{closed-shell active orbital 3}}
		\end{subfigure}

		\caption{\rev{\textbf{Active space orbitals visualization of indole rings.} Top row: (3e,3o) open shell active space orbitals; Bottom row: (4e,3o) closed shell active space orbitals. These orbitals are Hartree-Fock canonical orbitals obtained by PySCF's CASCI module \cite{sun2020}. The yellow marked atoms are the hydrogen link-atoms. The Isosurfaces are rendered at value 0.03 a.u. using ChimeraX \cite{meng2023}.}}
		\label{fig:active_space_orbitals}
	\end{figure}
	
	\section{Supplementary Tables}
	
	% REV: 整个表格改动，针对 referee 1, comment 5
	\begin{table}[H]
	\begin{revblock}
		\centering
		\begin{adjustbox}{center}
			\begin{tabular}{|c|c|c|c|cccc|c|}
				\hline
				\multirow{2}{*}{\textbf{system}} &
				\multirow{2}{*}{\textbf{ansatz}} &
				\multirow{2}{*}{\textbf{\begin{tabular}[c]{@{}c@{}}active\\ space\end{tabular}}} &
				\multirow{2}{*}{\textbf{qubits}} &
				\multicolumn{4}{c|}{\textbf{logical circuit}} &
				\multirow{2}{*}{\textbf{\begin{tabular}[c]{@{}c@{}}Pauli\\ words \\(truncated)\end{tabular}}} \\ \cline{5-8}
				&
				&
				&
				&
				\multicolumn{1}{c|}{\textbf{\begin{tabular}[c]{@{}c@{}}circuit\\ depth\end{tabular}}} &
				\multicolumn{1}{c|}{\textbf{\begin{tabular}[c]{@{}c@{}}total\\ gates\end{tabular}}} &
				\multicolumn{1}{c|}{\textbf{\begin{tabular}[c]{@{}c@{}}1-qubit\\ gates\end{tabular}}} &
				\textbf{\begin{tabular}[c]{@{}c@{}}2-qubit\\ gates\end{tabular}} &
				\\ \hline
				$NH_3\cdots FCl$    & \multirow{15}{*}{UCCSD} & (10e,10o) & 20 & \multicolumn{1}{c|}{9164}  & \multicolumn{1}{c|}{14079} & \multicolumn{1}{c|}{671} & 13408 & 6389  \\ \cline{1-1} \cline{3-9} 
				$NH_3\cdots Cl_2$   &                         & (10e,10o) & 20 & \multicolumn{1}{c|}{9354}  & \multicolumn{1}{c|}{14425} & \multicolumn{1}{c|}{677} & 13748 & 6107  \\ \cline{1-1} \cline{3-9} 
				$NH_3\cdots F_2$    &                         & (10e,10o) & 20 & \multicolumn{1}{c|}{7220}  & \multicolumn{1}{c|}{11251} & \multicolumn{1}{c|}{539} & 10712 & 6157  \\ \cline{1-1} \cline{3-9} 
				$HCN\cdots FCl$     &                         & (8e,8o)   & 16 & \multicolumn{1}{c|}{1592}  & \multicolumn{1}{c|}{2628}  & \multicolumn{1}{c|}{140} & 2488  & 2817  \\ \cline{1-1} \cline{3-9} 
				$H_2O\cdots FCl$    &                         & (10e,10o) & 20 & \multicolumn{1}{c|}{10987} & \multicolumn{1}{c|}{17135} & \multicolumn{1}{c|}{835} & 16300 & 12647 \\ \cline{1-1} \cline{3-9} 
				$C_2H_2\cdots FCl$  &                         & (10e,10o) & 20 & \multicolumn{1}{c|}{3014}  & \multicolumn{1}{c|}{4825}  & \multicolumn{1}{c|}{217} & 4608  & 3727  \\ \cline{1-1} \cline{3-9} 
				$C_2H_4\cdots F_2$  &                         & (10e,10o) & 20 & \multicolumn{1}{c|}{10949} & \multicolumn{1}{c|}{17085} & \multicolumn{1}{c|}{833} & 16252 & 4645  \\ \cline{1-1} \cline{3-9} 
				$NH_3$              &                         & (8e,8o)   & 16 & \multicolumn{1}{c|}{2711}  & \multicolumn{1}{c|}{4436}  & \multicolumn{1}{c|}{252} & 4184  & 2461  \\ \cline{1-1} \cline{3-9} 
				$HCN$               &                         & (6e,6o)   & 12 & \multicolumn{1}{c|}{388}   & \multicolumn{1}{c|}{689}   & \multicolumn{1}{c|}{45}  & 644   & 551   \\ \cline{1-1} \cline{3-9} 
				$H_2O$              &                         & (8e,8o)   & 16 & \multicolumn{1}{c|}{1984}  & \multicolumn{1}{c|}{3260}  & \multicolumn{1}{c|}{180} & 3080  & 2305  \\ \cline{1-1} \cline{3-9} 
				$C_2H_2$            &                         & (10e,10o) & 20 & \multicolumn{1}{c|}{3255}  & \multicolumn{1}{c|}{5229}  & \multicolumn{1}{c|}{241} & 4988  & 4007  \\ \cline{1-1} \cline{3-9} 
				$C_2H_4$            &                         & (10e,10o) & 20 & \multicolumn{1}{c|}{9344}  & \multicolumn{1}{c|}{14633} & \multicolumn{1}{c|}{709} & 13924 & 4159  \\ \cline{1-1} \cline{3-9} 
				$FCl$               &                         & (2e,2o)   & 4  & \multicolumn{1}{c|}{9}     & \multicolumn{1}{c|}{15}    & \multicolumn{1}{c|}{3}   & 12    & 15    \\ \cline{1-1} \cline{3-9} 
				$Cl_2$              &                         & (2e,2o)   & 4  & \multicolumn{1}{c|}{9}     & \multicolumn{1}{c|}{15}    & \multicolumn{1}{c|}{3}   & 12    & 15    \\ \cline{1-1} \cline{3-9} 
				$F_2$               &                         & (2e,2o)   & 4  & \multicolumn{1}{c|}{9}     & \multicolumn{1}{c|}{15}    & \multicolumn{1}{c|}{3}   & 12    & 15    \\ \hline
			\end{tabular}
		\end{adjustbox}
		\caption{\rev{\textbf{Quantum resource summary for calculating dissociation energies of Charge-Transfer dataset.} The first 7 rows are dimers in CT7/04 dataset, while the other rows are monomers of the above dimers. For the largest CT dimer, $C_2H_4\cdots F_2$, we employed an active space of (10e,10o), which is slightly smaller than the (14e,14o) space used in the original MC-PDFT study, in order to satisfy the limited simulated qubit count within 20 qubits for this proof-of-concept VQE-PDFT demonstration. The reported `Pauli words' counts correspond to the Pauli decomposition of the CASCI active-space Hamiltonian after applying the default TenCirChem \cite{li2023c} coefficient cutoff (discarding terms with $|c|<10^{-12}$) prior to fermion-to-qubit mapping (Jordan-Wigner transformation here). Note that all calculations were conducted by noiseless simulation rather than any actual quantum hardware.}}
	\end{revblock}
	\end{table}
	
	% REV: 新增了整个表格，针对 referee 1, comment 5
	\begin{table}[H]
	\begin{revblock}
		\centering
		\begin{subtable}[t]{0.9\textwidth}
			\centering
			\caption{\textbf{ROUCCSD and HEA resource summary}}
				\begin{adjustbox}{center}
					\begin{tabular}{|c|c|c|c|cccc|cc|}
						\hline
						\multirow{2}{*}{\textbf{\begin{tabular}[c]{@{}c@{}}system\\ ErCRY4\end{tabular}}} &
						\multirow{2}{*}{\textbf{ansatz}} &
						\multirow{2}{*}{\textbf{\begin{tabular}[c]{@{}c@{}}active\\ space\end{tabular}}} &
						\multirow{2}{*}{\textbf{qubits}} &
						\multicolumn{4}{c|}{\textbf{logical circuit}} &
						\multicolumn{2}{c|}{\textbf{Pauli words}} \\ \cline{5-10} 
						&
						&
						&
						&
						\multicolumn{1}{c|}{\textbf{\begin{tabular}[c]{@{}c@{}}circuit\\ depth\end{tabular}}} &
						\multicolumn{1}{c|}{\textbf{\begin{tabular}[c]{@{}c@{}}total\\ gates\end{tabular}}} &
						\multicolumn{1}{c|}{\textbf{\begin{tabular}[c]{@{}c@{}}1-qubit\\ gates\end{tabular}}} &
						\textbf{\begin{tabular}[c]{@{}c@{}}2-qubit\\ gates\end{tabular}} &
						\multicolumn{1}{c|}{\textbf{truncated}} &
						\textbf{total} \\ \hline
						\multirow{2}{*}{open shell}   & ROUCCSD & (3e,3o) & 4 & \multicolumn{1}{c|}{35} & \multicolumn{1}{c|}{55} & \multicolumn{1}{c|}{7} & 48 & \multicolumn{1}{c|}{118} & 262 \\ \cline{2-10} 
						                              & OHEA    & (3e,3o) & 4 & \multicolumn{1}{c|}{6}  & \multicolumn{1}{c|}{7}  & \multicolumn{1}{c|}{2} & 5  & \multicolumn{1}{c|}{100} & 219 \\ \hline
						\multirow{2}{*}{closed shell} & ROUCCSD & (4e,3o) & 4 & \multicolumn{1}{c|}{35} & \multicolumn{1}{c|}{56} & \multicolumn{1}{c|}{8} & 48 & \multicolumn{1}{c|}{118} & 262 \\ \cline{2-10} 
						                              & CHEA    & (4e,3o) & 4 & \multicolumn{1}{c|}{4}  & \multicolumn{1}{c|}{5}  & \multicolumn{1}{c|}{2} & 3  & \multicolumn{1}{c|}{100} & 219 \\ \hline
					\end{tabular}
				\end{adjustbox}
		\end{subtable}
		
		\vspace{6pt}
		
		\begin{subtable}[t]{0.9\textwidth}
			\centering
			\caption{\textbf{HEA resource summary after compiling to a 13-qubit chip}}
			\begin{adjustbox}{center}
				\begin{tabular}{|l|c|c|c|cccc|cc|}
					\hline
					\multicolumn{1}{|c|}{\multirow{2}{*}{\textbf{\begin{tabular}[c]{@{}c@{}}system\\ ErCRY4\end{tabular}}}} &
					\multirow{2}{*}{\textbf{ansatz}} &
					\multirow{2}{*}{\textbf{\begin{tabular}[c]{@{}c@{}}active\\ space\end{tabular}}} &
					\multirow{2}{*}{\textbf{qubits}} &
					\multicolumn{4}{c|}{\textbf{transpiled circuit}} &
					\multicolumn{2}{c|}{\textbf{Commuting groups}} \\ \cline{5-10} 
					\multicolumn{1}{|c|}{} &
					&
					&
					&
					\multicolumn{1}{c|}{\textbf{\begin{tabular}[c]{@{}c@{}}circuit\\ depth\end{tabular}}} &
					\multicolumn{1}{c|}{\textbf{\begin{tabular}[c]{@{}c@{}}total\\ gates\end{tabular}}} &
					\multicolumn{1}{c|}{\textbf{\begin{tabular}[c]{@{}c@{}}1-qubit\\ gates\end{tabular}}} &
					\textbf{\begin{tabular}[c]{@{}c@{}}2-qubit\\ gates\end{tabular}} &
					\multicolumn{1}{c|}{\textbf{\begin{tabular}[c]{@{}c@{}}global\\ (Hamiltonian)\end{tabular}}} &
					\textbf{\begin{tabular}[c]{@{}c@{}}local\\ (RDMs)\end{tabular}} \\ \hline
					open shell &
					OHEA &
					(3e,3o) &
					4 &
					\multicolumn{1}{c|}{46} &
					\multicolumn{1}{c|}{61} &
					\multicolumn{1}{c|}{46} &
					15 &
					\multicolumn{1}{c|}{\multirow{2}{*}{26}} &
					\multirow{2}{*}{110} \\ \cline{1-8}
					closed shell &
					CHEA &
					(4e,3o) &
					4 &
					\multicolumn{1}{c|}{18} &
					\multicolumn{1}{c|}{23} &
					\multicolumn{1}{c|}{16} &
					7 &
					\multicolumn{1}{c|}{} &
					\\ \hline
				\end{tabular}
			\end{adjustbox}
		\end{subtable}
		\caption{\rev{\textbf{Quantum resource summary for the active space of QM region in ErCRY4 protein.} (a) In computation with ROUCCSD, the Pauli words were generated by Jordan-Wigner transformation, while the HEA circuits employed parity transformation with symmetry tapering (two qubits) in the fixed $(N_\alpha,N_\beta)$ sector, thereby reducing the number of distinct Pauli terms to be measured. The “truncated” denotes Pauli words after the default TenCirChem coefficient cutoff ($|c|<10^{-12}$), whereas “total” reports the corresponding untruncated term counts. (b) After compilation process, the circuit depth of HEAs increased compared to the logical circuit since the logical gates were decomposed by native gate set. Pauli terms were partitioned into qubit-wise commuting (QWC) groups using the built-in TensorCircuit routine \cite{zhang2023b}, such that any two operators in the same group never act with different non-identity Pauli operators on the same qubit, therefore may be measured simultaneously ($100 \to 26$ global grouping for the truncated Hamiltonian evaluation at each step of VQE). On the other hand, the RDM measurements were performed matrix-element-wise (`total') and thus employ only element-wise QWC grouping ($219 \to 110$ local groups for the full RDM operator set).}}
	\end{revblock}
	\end{table}
	
	% REV: 新增了整个表格，针对 referee 1, comment 8
	\begin{table}[H]
	\begin{revblock}
		\centering
		\begin{tabular}{|c|l|l|l|l|}
			\hline
			\multicolumn{1}{|l|}{\textbf{Dimer}} &
			\textbf{Methods} &
			\textbf{\begin{tabular}[c]{@{}l@{}}E Dimer\\ (Hartree)\end{tabular}} &
			\textbf{\begin{tabular}[c]{@{}l@{}}E m-1\\ (Hartree)\end{tabular}} &
			\textbf{\begin{tabular}[c]{@{}l@{}}E m-2\\ (Hartree)\end{tabular}} \\ \hline \hline
			\multirow{3}{*}{$NH_3\cdots FCl$}   & VQE      & -615.128804 & -56.221505 & -558.898720 \\ \cline{2-5} 
			& VQE-PDFT & -616.242235 & -56.503348 & -559.720469 \\ \cline{2-5} 
			& MC-PDFT  & -616.223810 & -56.490022 & -559.714003 \\ \hline\hline
			\multirow{3}{*}{$NH_3\cdots Cl_2$}  & VQE      & -975.223380 & -56.221467 & -918.999406 \\ \cline{2-5} 
			& VQE-PDFT & -976.527014 & -56.503455 & -920.015446 \\ \cline{2-5} 
			& MC-PDFT  & -976.507738 & -56.490022 & -920.010464 \\ \hline\hline
			\multirow{3}{*}{$NH_3\cdots F_2$}   & VQE      & -254.974775 & -56.221427 & -198.753481 \\ \cline{2-5} 
			& VQE-PDFT & -255.898334 & -56.503449 & -199.391897 \\ \cline{2-5} 
			& MC-PDFT  & -255.867282 & -56.490022 & -199.375496 \\ \hline\hline
			\multirow{3}{*}{$HCN\cdots FCl$}    & VQE      & -651.815085 & -92.908656 & -558.902946 \\ \cline{2-5} 
			& VQE-PDFT & -653.055775 & -93.329282 & -559.721513 \\ \cline{2-5} 
			& MC-PDFT  & -653.034226 & -93.314369 & -559.714003 \\ \hline\hline
			\multirow{3}{*}{$H_2O\cdots FCl$}   & VQE      & -634.982061 & -76.063217 & -558.902897 \\ \cline{2-5} 
			& VQE-PDFT & -636.098589 & -76.371825 & -559.721511 \\ \cline{2-5} 
			& MC-PDFT  & -636.070153 & -76.349168 & -559.714003 \\ \hline\hline
			\multirow{3}{*}{$C_2H_2\cdots FCl$} & VQE      & -635.759503 & -76.850009 & -558.902915 \\ \cline{2-5} 
			& VQE-PDFT & -636.962657 & -77.236253 & -559.721512 \\ \cline{2-5} 
			& MC-PDFT  & -636.948390 & -77.227991 & -559.714003 \\ \hline\hline
			\multirow{3}{*}{$C_2H_4\cdots F_2$} & VQE      & -276.817798 & -78.065357 & -198.754252 \\ \cline{2-5} 
			& VQE-PDFT & -277.876723 & -78.483346 & -199.392029 \\ \cline{2-5} 
			& MC-PDFT  & -277.852830 & -78.476804 & -199.375496 \\ \hline
		\end{tabular}
		\caption{\rev{\textbf{The absolute energies of dimers and monomers in the Charge-Transfer dataset.} The results were obtained via VQE in CASCI level, VQE-PDFT, and the MC-PDFT method, while the MC-PDFT's results were directly fetched from the Supplementary Material of Ref. \cite{ghosh2015}. Note that the VQE-PDFT energies are based on CASCI framework, whereas the literature MC-PDFT values rely on CASSCF framework. Because MC-PDFT is not variational with respect to the choice of multiconfigurational reference, these absolute energies are provided for general reference rather than for ranking the methods. For assessing accuracy in dissociation energies, the MUE values reported in Table 1 are more informative. Here, all calculations were performed in the jul-cc-pVTZ basis \cite{papajak2011}. The tPBE functional was adopted in both VQE-PDFT and MC-PDFT calculations. The m1 and m2 refer to the two monomers in the m1-m2 dimer, respectively.}}
	\end{revblock}
	\end{table}
	
	\begin{table}[H]
		\centering
		\begin{tabular}{|c|c|c|}
			\hline
			\textbf{frame} & $\Delta G^o$ & $\lambda$  \\ \hline
			157            & 0.099934408             & 0.549241928     \\ \hline
			264            & 0.043676293             & 0.347217505     \\ \hline
			461            & 0.040885297             & 0.220829577     \\ \hline
			631            & 0.163803428             & 0.517004701     \\ \hline
			682            & 0.038024409             & 0.495626829     \\ \hline
			706            & 0.071676358             & 0.307290925     \\ \hline
			777            & 0.037682775             & 0.547830361     \\ \hline
			1077           & 0.033809862             & 0.53029316      \\ \hline
			1107           & 0.167015044             & 0.532070337     \\ \hline
			1237           & 0.084752646             & 0.131341481     \\ \hline
			1243           & 0.023647298             & 0.623133927     \\ \hline
			1311           & 0.025645922             & 0.415820148     \\ \hline
			1368           & 0.05873396              & 0.774978721     \\ \hline
			1502           & 0.151100112             & 0.464021283     \\ \hline
			1603           & 0.113313698             & 0.217739833     \\ \hline
			1626           & 0.000816062             & 0.288449438     \\ \hline
			1627           & 0.021599893             & 0.345962916     \\ \hline
			1965           & 0.115245417             & 0.488976157     \\ \hline
			2071           & 0.13571487              & 0.340026962     \\ \hline
			2180           & 0.020222078             & 0.57379004      \\ \hline
		\end{tabular}
		\caption{\textbf{$\Delta G^o$ and $\lambda$ calculations by empirical HEA on noiseless simulators.} The 20 protein conformations were randomly sampled from the MD simulation of GS ErCRY4 protein in reference \cite{xu2021}. The frame numbers indicate locations of conformations during the entire simulation process. All results are given in eV.}
	\end{table}
	
	\begin{table}[H]
		\centering
		\begin{tabular}{|c|c|c|}
			\hline
			\textbf{frame} & $H_{DA}$ & $|H_{DA}|^2$  \\ \hline
			157            & -9.44011E-04                       & 8.91158E-07                          \\ \hline
			264            & 1.50656E-02                        & 2.26972E-04                          \\ \hline
			461            & -3.43295E-02                       & 1.17851E-03                          \\ \hline
			631            & -2.49168E-04                       & 6.20846E-08                          \\ \hline
			682            & -2.21206E-03                       & 4.89319E-06                          \\ \hline
			706            & -1.65424E-04                       & 2.73650E-08                          \\ \hline
			777            & -4.04790E-04                       & 1.63855E-07                          \\ \hline
			1077           & 1.48556E-02                        & 2.20688E-04                          \\ \hline
			1107           & -3.80092E-04                       & 1.44470E-07                          \\ \hline
			1237           & 1.37738E-03                        & 1.89717E-06                          \\ \hline
			1243           & 1.67444E-03                        & 2.80376E-06                          \\ \hline
			1311           & 5.25197E-03                        & 2.75831E-05                          \\ \hline
			1368           & -1.37476E-03                       & 1.88996E-06                          \\ \hline
			1502           & -1.84872E-02                       & 3.41775E-04                          \\ \hline
			1603           & 3.46027E-03                        & 1.19735E-05                          \\ \hline
			1626           & 2.59327E-03                        & 6.72507E-06                          \\ \hline
			1627           & 3.57321E-03                        & 1.27678E-05                          \\ \hline
			1965           & -1.51863E-02                       & 2.30625E-04                          \\ \hline
			2071           & -2.02746E-03                       & 4.11059E-06                          \\ \hline
			2180           & 3.41750E-03                        & 1.16793E-05                          \\ \hline
		\end{tabular}
		\caption{\textbf{Electronic coupling calculations.} The 6-31G(d) basis set and the CAM-B3LYP functional were used in all calculations. The 20 conformations were randomly sampled from the MD simulation of GS ErCRY4 protein in reference \cite{xu2021}. The frame numbers indicate locations of conformations during the entire simulation process. All results are given in eV.}
	\end{table}
	
	\begin{table}[H]
		\centering
		\begin{tabular}{|c|c|c|}
			\hline
			\textbf{frame} & $\Delta G^o$ & $\lambda$  \\ \hline
			157            & 0.075376004                        & 0.533793409                          \\ \hline
			264            & 0.066947358                        & 0.550103409                          \\ \hline
			461            & 0.022934202                        & 0.529776474                          \\ \hline
			631            & 0.098816411                        & 0.654202407                          \\ \hline
			682            & 0.102475986                        & 0.552455689                          \\ \hline
			706            & 0.109047811                        & 0.593777179                          \\ \hline
			777            & 0.091658923                        & 0.600336874                          \\ \hline
			1077           & 0.081086545                        & 0.581411027                          \\ \hline
			1107           & 0.082290769                        & 0.627811309                          \\ \hline
			1237           & 0.039260208                        & 0.569951056                          \\ \hline
			1243           & 0.110907047                        & 0.655571383                          \\ \hline
			1311           & 0.004307911                        & 0.521179077                          \\ \hline
			1368           & 0.035574098                        & 0.544496837                          \\ \hline
			1502           & 0.131879075                        & 0.603524833                          \\ \hline
			1603           & 0.003378448                        & 0.516846946                          \\ \hline
			1626           & 0.068215042                        & 0.512984809                          \\ \hline
			1627           & 0.036969471                        & 0.529376022                          \\ \hline
			1965           & 0.123957956                        & 0.655946507                          \\ \hline
			2071           & 0.043474286                        & 0.524336485                          \\ \hline
			2180           & 0.049975887                        & 0.543812624                          \\ \hline
		\end{tabular}
		\caption{\textbf{$\Delta G^o$ and $\lambda$ calculations by ROUCCSD on noiseless simulators.} The 20 protein conformations were randomly sampled from the MD simulation of GS ErCRY4 protein in reference \cite{xu2021}. The frame numbers indicate locations of conformations during the entire simulation process. All results are given in eV.}
	\end{table}
	
	\begin{table}[H]
		\centering
		\begin{tabular}{|ccc|c|c|}
			\hline
			\multicolumn{1}{|c|}{\textbf{index}} & \multicolumn{1}{c|}{\textbf{system}}                 & \textbf{shell type} & \textbf{\begin{tabular}[c]{@{}c@{}}Empirical HEA\\ noiseless\end{tabular}} & \textbf{ROUCCSD} \\ \hline
			\multicolumn{1}{|c|}{1}     & \multicolumn{1}{c|}{point-a TrpB\textsuperscript{+}} & open-shell          & -362.99415                                                                 & -362.99942       \\ \hline
			\multicolumn{1}{|c|}{2}     & \multicolumn{1}{c|}{point-a TrpC\textsuperscript{0}} & closed-shell        & -363.26459                                                                 & -363.26554       \\ \hline
			\multicolumn{1}{|c|}{3}     & \multicolumn{1}{c|}{point-b TrpB\textsuperscript{0}} & closed-shell        & -363.26777                                                                 & -363.26682       \\ \hline
			\multicolumn{1}{|c|}{4}     & \multicolumn{1}{c|}{point-b TrpC\textsuperscript{+}} & open-shell          & -362.98880                                                                 & -362.98758       \\ \hline
			\multicolumn{1}{|c|}{5}     & \multicolumn{1}{c|}{point-c TrpB\textsuperscript{0}} & closed-shell        & -363.26994                                                                 & -363.26920       \\ \hline
			\multicolumn{1}{|c|}{6}     & \multicolumn{1}{c|}{point-c TrpC\textsuperscript{+}} & open-shell          & -362.99378                                                                 & -362.99736       \\ \hline
			\multicolumn{1}{|c|}{7}     & \multicolumn{1}{c|}{point-d TrpB\textsuperscript{+}} & open-shell          & -362.99775                                                                 & -362.99268       \\ \hline
			\multicolumn{1}{|c|}{8}     & \multicolumn{1}{c|}{point-d TrpC\textsuperscript{0}} & closed-shell        & -363.26632                                                                 & -363.26518       \\ \hline\hline
			\multicolumn{3}{|c|}{$\lambda$ ($eV$)}                                                                   & 0.34003                                                                    & 0.52434          \\ \hline
			\multicolumn{3}{|c|}{$\Delta G^o$ ($eV$)}                                                                & 0.13571                                                                    & 0.04347          \\ \hline
			\multicolumn{3}{|c|}{$k_{ET}$ ($10^8 s^{-1}$)}                                                           & 1.89064                                                                    & 2.48679          \\ \hline
		\end{tabular}
		\caption{\textbf{Single-point energies, Marcus parameters, and the transfer rate of the random conformation.} The conformation was randomly selected as the 2071 frame from the 20 conformations above and calculated within the QM/MM framework. The calculations were conducted by ROUCCSD and empirical HEA on noiseless simulator. The single-point energies are given in Hartree. The 6-31G basis set and the tPBE functional were used in all calculations.}
	\end{table}
	
\bibliographystyle{vancouver}
\bibliography{ref}